\documentclass[12pt]{article}
\usepackage[utf8]{inputenc}
\usepackage[letterpaper, margin=1in]{geometry} % 8.5x11 with 1-inch margins
\usepackage{setspace}
\usepackage{algorithm}
\usepackage{algpseudocode}

\usepackage[dvipsnames,table,xcdraw]{xcolor}
\usepackage{graphicx}
\usepackage{xr}
\usepackage{hyperref}
\usepackage{subcaption}
\usepackage{mathtools}
\usepackage[title]{appendix}
\usepackage{bm}
\usepackage{natbib}
\usepackage{etoolbox}
\usepackage{enumitem}
\usepackage{amsmath}
\usepackage{amssymb}
\usepackage{amsthm}
\usepackage{bm}
\usepackage{bbold}
\usepackage[normalem]{ulem}
\usepackage[labelfont=bf]{caption}
\usepackage{tabularx}
\usepackage{booktabs}
\usepackage{adjustbox}
\usepackage{multirow}

\newcommand{\bs}{\boldsymbol}

\usepackage[normalem]{ulem}
\title{\vspace{-1cm}Similarity-Driven Proposals for MCMC Algorithms on Discrete Spaces}

\author{
  Luca Aiello\thanks{Department of Biostatistics, University of California, Los Angeles \href{mailto:laiello@g.ucla.edu}{laiello@g.ucla.edu}},
  Raffaele Argiento\thanks{Department of Economics, University of Bergamo, Italy},  
  Alexandros Beskos\thanks{Department of Statistical Science, University College of London, United Kingdom},  
  Maria De Iorio\thanks{Yong Loo Lin School of Medicine, National University of Singapore, Singapore}
}
\date{}

\begin{document}

\maketitle
\vspace{-1cm}
\begin{abstract}
    Recent research has led to the development of MCMC algorithms with likelihood-informed proposals when targeting posterior distributions supported on discrete state spaces. Our work is placed within this field and puts forward a new MCMC methodology based upon \emph{similarity-driven proposals}. Such  proposals sway transitions towards states favored by the posterior via use of a data-driven measure of discrepancy between observations and the proposed model. Our approach can naturally cover classes of hierarchical models that involve both discrete variables and additional latent ones, without a requirement of integrating our the latter, in contrast to previous works in this field. The new algorithms are illustrated in simulation settings and in a involved real data scenario with  a Dirichlet-Multinomial regression model.
\end{abstract}

\noindent%

%\vspace{-0.5cm}

\section{Introduction}

Efficient exploration of complex probability distributions is a central problem in computational statistics and Bayesian inference. Markov chain Monte Carlo (MCMC) methods provide a general framework for posterior approximation, uncertainty quantification and model selection when analytic solutions are unavailable. Their practical efficiency, however, depends critically on the choice of proposal distributions: poorly designed proposals can lead to slow mixing, high autocorrelation and ineffective exploration, especially in high-dimensional settings.

In continuous parameter spaces, informed MCMC methods that exploit structural information about the target distribution can lead to substantial efficiency gains. Gradient-based approaches such as the Metropolis-adjusted Langevin algorithm \citep[MALA;][]{roberts1998optimal} and Hamiltonian Monte Carlo \citep[HMC;][]{neal2011mcmc, girolami2011riemann} construct proposals aligned with the local geometry of the posterior, enabling larger moves with high acceptance probability. Extensions including Riemann manifold HMC, stochastic-gradient MCMC and accelerated Langevin schemes have further expanded these ideas to large-scale and non-conjugate models, demonstrating the benefits of incorporating even partial information about the target distribution \citep{welling2011bayesian, durmus2017fast, titsias2017hamming}.

Discrete state spaces pose qualitatively different challenges. In problems such as clustering, graphical model inference and Bayesian variable selection, the parameter space is combinatorial and lacks a natural notion of gradient, limiting the applicability of continuous-space techniques. Standard local proposals, such as single-component updates or random reassignments, often struggle to traverse the posterior efficiently, particularly in the presence of strong dependence, hierarchical structure or collinearity among variables.

MCMC strategies using likelihood-informed proposals have recently been developed for discrete spaces, highlighting the benefits of incorporating partial, computationally tractable information about the target distribution. The Hamming Ball sampler \citep{titsias2017hamming} restricts exploration to neighbourhoods defined by a bounded Hamming distance from the current state, enabling informed local moves while controlling computational cost in exponentially large state spaces. Locally balanced proposals \citep{zanella2020informed}  improve efficiency by reweighting local transition kernels using functions of the target density ratio, favouring moves toward higher-probability states while preserving detailed balance. Extending these ideas, informed reversible-jump MCMC \citep{gagnon2021informed} applies similar principles in trans-dimensional settings, using large-sample approximations of posterior model probabilities to guide proposals. Together, these methods demonstrate that even partial or approximate information about the target can substantially improve mixing and exploration in complex discrete inference problems.

Motivated by recent advances, we propose a new MCMC methodology for posteriors on discrete state spaces based on similarity-driven proposals. Unlike existing approaches \citep{zanella2020informed,gagnon2021informed}, our method does not require integrating out latent variables or access to closed-form marginal likelihoods. Instead, transitions are biased using a data-driven measure of discrepancy between observed data and candidate model fits, favouring states that better explain the data while preserving exact Metropolis-Hastings (MH) validity. By combining structured local moves with empirical similarity information, the sampler more effectively explores high-probability regions, improving mixing and acceptance rates. Our framework is broadly applicable across discrete inference problems, including clustering, graphical models and Bayesian variable selection, thus enabling efficient exploration of complex and high-dimensional model spaces.

Overall, this work helps bridging the gap between informed MCMC methods developed for continuous spaces and the demands of discrete combinatorial inference. By exploiting both local structural constraints and empirical similarity measures, we provide a flexible and theoretically sound framework for accelerating MCMC in high-dimensional discrete settings.

The remainder of the paper is organized as follows. Section~\ref{sec:smlrt_drvn_MH} introduces the proposed similarity-driven MH proposals and discusses main properties. Section~\ref{sec:simulations} presents simulation studies assessing performance in controlled scenarios. Sections~\ref{sec:vs_dm}~and~\ref{sec:rda} illustrate the methodology through variable selection in Dirichlet-Multinomial regression and a real-data application, respectively. Section~\ref{sec:disc} concludes with a discussion and directions for \mbox{future work.}

\section{Similarity-Driven Metropolis-Hastings}\label{sec:smlrt_drvn_MH}

\subsection{Our Approach}

We consider hierarchical Bayesian models with a discrete parameter $\bs{\xi} \in \Xi$ and additional continuous or latent parameters $\bs{\theta} \in \Theta$. The joint posterior is
\begin{equation}\label{eq:general}
    \pi(\bs{\theta}, \bs{\xi} \mid \bs{y})
    \propto p(\bs{y} \mid \bs{\theta}, \bs{\xi})\, p(\bs{\theta} \mid \bs{\xi})\, p(\bs{\xi})
\end{equation}
where $\bs{y}$ are the observed data. This general formulation encompasses a wide range of applications, including clustering models, graphical models and Bayesian variable selection. In graphical models, $\bs{\xi}$ encodes the graph structure in a space $\Xi \subseteq \{0,1\}^{|E|}$, where $E$ is the set of edges and each component indicates edge inclusion under constraints such as sparsity, decomposability or acyclicity. In clustering, $\bs{\xi}$ represents a partition of $\{1,\ldots,n\}$, equivalently described by cluster-label vectors up to permutation, forming a highly non-Euclidean combinatorial space. In variable selection, $\bs{\xi} \in \{0,1\}^P$ indexes subsets of $P$ predictors, with $|\Xi|$ growing exponentially in $P$. In each case, candidate states $\bs{\xi}$ induce model-implied summaries or predictive quantities that can be meaningfully compared to the observed data. This structure provides a natural foundation for the similarity-driven proposal mechanisms developed in the remainder of this section. 

Given a target distribution $\pi(\bs{\xi}\mid -)$ defined on a state space $\Xi$, MCMC methods simulate a Markov chain $\{\bs{\xi}_t\}_{t\ge 1}$ with stationary distribution $\pi$, using the visited states as Monte Carlo samples. Under mild regularity conditions, the ergodic theorem guarantees that sample averages converge to expectations under $\pi$. Many practical MCMC schemes are instances of the MH algorithm \citep{metropolis1953equation,hastings1970monte}. From a current state $\bs{\xi} \in \Xi$, MH proposes a candidate $\bs{\xi}^\prime \sim Q(\cdot \mid \bs{\xi})$ and accepts it with probability
\begin{equation*}
    \alpha(\bs{\xi},\bs{\xi}^\prime) = \min\left\{1, \frac{\pi(\bs{\xi}^\prime\mid -)\,Q(\bs{\xi}\mid\bs{\xi}^\prime)}{\pi(\bs{\xi} \mid - )\,Q(\bs{\xi}^\prime\mid\bs{\xi})}\right\}
\end{equation*}
otherwise the chain remains at $\bs{\xi}$. While the MH framework is general, efficiency depends critically on the choice of the proposal $Q$, with poor choices resulting in slow mixing and convergence, while well-designed proposals improve upon both of these chain properties.

Formally, let $\pi$ be a target distribution on a discrete state space $\Xi$, and let $K(\bs{\xi}^\prime \mid \bs{\xi})$ denote a baseline uninformed local proposal (e.g., the uniform distribution) supported on a neighborhood $N(\bs{\xi})$ (e.g., the set of configurations considered ``close'' to $\bs{\xi}$). A simple strategy to inform $K$ is to reweight candidate moves according to relative posterior probabilities. \citet{zanella2020informed} introduce \emph{pointwise informed proposals} of the form
\begin{equation*}
    Q_g(\bs{\xi}^\prime \mid \bs{\xi}) \propto 
    g\left(\frac{\pi(\bs{\xi}^\prime \mid \bs{y})}{\pi(\bs{\xi}\mid \bs{y})}\right) \,
    K(\bs{\xi}^\prime \mid \bs{\xi}) \, \mathbf{1}(\bs{\xi}^\prime \in N(\bs{\xi}))
\end{equation*}
where $g:(0,\infty)\to(0,\infty)$ controls the bias toward higher-probability states with the proportionality hiding the normalization constant. Choosing $g\equiv 1$ recovers the uninformed kernel, while other choices (e.g., $g(u)=\sqrt{u}$) yield locally balanced proposals and extend naturally to trans-dimensional moves \citep{gagnon2021informed}. While such constructions are theoretically appealing, their effectiveness may be limited in complex discrete spaces where posterior mass is highly irregular or concentrated, and where evaluating posterior ratios may be computationally challenging, motivating alternative ways of incorporating problem-specific structure.

We propose a \emph{similarity-driven proposal} that leverages the discrepancy between the observed data and a candidate state $\bs{\xi}^\prime$. We start by defining the unnormalised transition
\begin{equation}\label{eq:dis_prop} 
    g(\bs{\xi}^\prime \mid \bs{\xi})
    = \exp\Big\{\big[-d\big(\mathcal{S}(\bs{y}), \widehat{\mathcal{S}}(\bs{\xi}^\prime)\big)\big]^\lambda\Big\} \, \mathbf{1}(\bs{\xi}^\prime \in N(\bs{\xi}))
\end{equation}
where $\mathcal{S}(\bs{y})$ are observed summary statistics, $\widehat{\mathcal{S}}(\bs{\xi}^\prime)$ the corresponding model-implied summary, $d(\cdot,\cdot)$ a dissimilarity metric taking values in $(-\infty,0)$ for the particular choices put forward in this work, and $\lambda>0$ controls informativeness. Values of $d(\cdot,\cdot)$ close to $0$ point to high dissimilarity, while negative values far from $0$ indicate small dissimilarity.
As such $\exp\Big\{\big[-d\big(\mathcal{S}(\bs{y}), \widehat{\mathcal{S}}(\bs{\xi}^\prime)\big)\big]^\lambda\Big\}$ provides a measure of similarity between the data and the proposed model. 
Because $g$ is defined via a directly computable, data-driven discrepancy rather than the exact posterior, the resulting similarity-driven proposal kernel
\begin{equation}\label{eq:similarity_driven}
    Q_{\lambda}(\bs{\xi}^\prime \mid \bs{\xi}) = 
    \frac{\exp\left\{\big[-d(\mathcal{S}(\bs{y}), \widehat{\mathcal{S}}(\bs{\xi}^\prime))\big]^\lambda\right\}\, K(\bs{\xi}^\prime \mid \bs{\xi})}{Z(\bs{\xi})} \, \mathbf{1}(\bs{\xi}^\prime \in N(\bs{\xi}))
\end{equation}
is fully computable, allowing exact MH updates without requiring marginal likelihoods. Here, $K(\bs{\xi}^\prime \mid \bs{\xi})$ is an uninformed local kernel and $Z(\bs{\xi})$ the normalization constant. As long as $Q_\lambda$ is a proper probability kernel, the chain targets the true posterior $\pi(\bs{\xi})$ favoring moves toward states better aligned with the observed data while avoiding the complications of other informed discrete-space proposals \citep{zanella2020informed,gagnon2021informed}. 

The choice of the dissimilarity measure $d(\cdot,\cdot)$ is deliberately left flexible and can be tailored to the structure of the latent state space. In particular, when $\bs{\xi}$ encodes combinatorial objects such as graphs or partitions, $d(\cdot,\cdot)$ may be chosen to reflect structural discrepancies, for instance through graph edit distances (i.e., the number of edge modifications needed to match two graphs), spectral distances, or partition-based criteria such as variation of information or Binder-type losses. This modularity allows the similarity-driven proposal to be adapted to a wide range of discrete and mixed discrete-continuous settings without altering the underlying sampling framework. In the context of variable selection, where $\bs{\xi}$ represents a vector of inclusion indicators, specific choices of dissimilarity measures and summary statistics are discussed in detail in Section~\ref{subsec:choices_sim}, together with their computational and inferential implications.

The tuning parameter $\lambda$ governs the balance between \emph{exploration} and \emph{exploitation} in the proposal mechanism. A small $\lambda$ leads to a nearly constant $g(\bs{\xi}^\prime \mid \bs{\xi})$, yielding proposals that approximate uninformed random-walk behavior and facilitate broad exploration of the state space. As $\lambda$ increases, the proposal distribution progressively concentrates around candidate states whose $\widehat{\mathcal{S}}(\bs{\xi}^\prime)$ closely resemble the observed data $\mathcal{S}(\bs{y})$, thereby promoting data-informed moves and faster convergence. However, overly large values of $\lambda$ may produce excessively concentrated proposals, leading to poor mixing and low acceptance probabilities as the chain becomes confined to local modes. Hence, $\lambda$ acts as a natural calibration parameter that controls the trade-off between global exploration and local refinement, influencing both the acceptance rate and overall sampling efficiency.

\subsection{Connections with other approaches}

In this subsection, we assume for simplicity that dissimilarity functions used in mathematical formulae take non-negative values. From a decision-theoretic perspective, $d(\mathcal{S}(\bs{y}), \widehat{\mathcal{S}}(\bs{\xi}^\prime))$ can be seen as a \emph{loss function} for selecting candidate model $\bs{\xi}^\prime$ \citep{berger2013statistical,robert2007bayesian,zellner1986bayesian}. Smaller values indicate models that better balance fit and complexity, incurring lower loss. The exponential weighting in \eqref{eq:dis_prop} implements a \emph{soft decision rule}, assigning higher proposal probabilities to lower-loss models while still allowing occasional moves to less favorable states. Alternative weighting schemes can also be considered, such as bounded kernels \citep[e.g., bisquare kernels,][]{ronchetti2009robust}, which may offer improved robustness or exploration in practice. This ensures stochastic exploration of the model space and systematically biases proposals toward models with improved explanatory power, analogous to a probabilistic version of classical utility-maximization.

In addition, the construction in \eqref{eq:dis_prop} has natural minimum-distance \citep{basu2011statistical} and generalized Bayesian \citep{bissiri2016general} interpretations. Exponentially weighting the discrepancy $d(\mathcal{S}(\bs{y}), \widehat{\mathcal{S}}(\bs{\xi}^\prime))$ parallels minimum-distance estimation, assessing models via a loss quantifying lack of fit. From a Bayesian viewpoint, the proposed construction can be interpreted through the lens of generalized Bayesian inference via a composite loss function. Specifically, consider 
\begin{equation*}
\mathcal{L}_\lambda(\bs{\xi}; \bs{y})
=
-\log p(\bs{y} \mid \bs{\xi})
+
d\left(\mathcal{S}(\bs{y}), \widehat{\mathcal{S}}(\bs{\xi})\right)^\lambda 
\end{equation*}
which augments the negative log-likelihood with a dissimilarity-based penalty measuring the mismatch between observed summaries and their model-implied counterparts. The resulting generalized Bayes posterior is proportional to
\begin{equation*}
p(\bs{\xi}) \, p(\bs{y} \mid \bs{\xi})
\exp\left\{
- d\left(\mathcal{S}(\bs{y}), \widehat{\mathcal{S}}(\bs{\xi})\right)^\lambda
\right\}
\end{equation*}

Within an MH algorithm, the acceptance probability 
 combines the usual likelihood ratio with an additional exponential distance-tilting factor, closely mirroring standard arguments that connect ABC to Gibbs posteriors. To clarify this connection, recall that the ABC posterior is commonly defined as
\begin{equation*}
\pi_{ABC}(\bs{\xi} \mid \bs{y})
\propto
p(\bs{\xi})
\int p(\bs{y}^* \mid \bs{\xi})
\, K_\lambda \left(
d\left(\mathcal{S}(\bs{y}), \mathcal{S}(\bs{y}^*)\right)
\right)
\, d\bs{y}^* 
\end{equation*}
where $\bs{y}^*$ denotes a dataset simulated from the model under $\bs{\xi}$, $\mathcal{S}(\cdot)$ is a vector of summary statistics and $K_\lambda$ is a kernel that assigns higher weight to simulated summaries close to observed ones \citep{jarvenpaa2025surrogate}. This formulation can be equivalently expressed as a generalized Bayes posterior
\begin{equation*}
\pi_{GB}(\bs{\xi} \mid \bs{y})
\propto
p(\bs{\xi}) \,
\mathbb{E}_{\bs{y}^* \mid \bs{\xi}}
\Big[
K_\lambda \left(
d \left(\mathcal{S}(\bs{y}), \mathcal{S}(\bs{y}^*)\right)
\right)
\Big]
\end{equation*}

In our similarity-driven proposal, we replace the stochastic summary $\mathcal{S}(\bs{y}^*)$ with the model-implied summary $\widehat{\mathcal{S}}(\bs{\xi})$, which can be interpreted as representative or expected summary under $p(\cdot \mid \bs{\xi})$. Moreover, the ABC kernel is replaced by an exponential discrepancy
\begin{equation*}
K_\lambda \left(
d\left(\mathcal{S}(\bs{y}), \mathcal{S}(\bs{y}^*)\right)
\right) =
\exp\left\{
- d\left(\mathcal{S}(\bs{y}), \widehat{\mathcal{S}}(\bs{\xi})\right)^\lambda
\right\}
\end{equation*}
yielding a generalized-Bayes-type weighting that is computationally efficient. {This construction is conceptually related to loss- or discrepancy-based generalized Bayesian approaches \citep{bissiri2016general} and, more recently, to theoretical developments on discrepancy-based posterior concentration \citep{legramanti2025concentration}. Crucially, this loss-based construction is (in our case) used solely to inform the proposal mechanism and the MH correction ensures that inference remains exact with respect to (w.r.t.)~the true posterior distribution. 

\subsection{Selection of the similarity function}\label{subsec:choices_sim}

The performance of our proposal in \eqref{eq:similarity_driven} depends critically on the choice of the dissimilarity function $d(\cdot \, ,\cdot)$. From the decision-theoretic perspective outlined above, $d(\cdot \, ,\cdot)$ plays the role of a loss function, quantifying the cost of selecting a candidate model $\bs{\xi}^\prime$ to explain the observations $\bs{y}$. In what follows, let $\widehat{\mathcal{S}}(\bs{\xi}^\prime)$ be the summary statistic evaluated under candidate model $\bs{\xi}^\prime$ and $\mathcal{S}(\bs{y})$ under the observed data. Different losses induce different controls of the balance between exploration and exploitation in the model space. 

Classical choices include distributional distances such as the Cramér-von Mises and Kolmogorov-Smirnov statistics \citep{cramer1928composition,smirnov1948table}, as well as broader classes of $\phi$-divergences and robust criteria \citep{basu2011statistical}. In stochastic model search, these quantities naturally translate into loss functions that favor candidates which achieve better empirical fit. In this section, we focus on test-based dissimilarities that connect the minimum-distance philosophy with familiar tools from classical model comparison. 

A natural way to quantify dissimilarity in model search is to compare a candidate model with a nested baseline and assess the improvement in fit. In classical settings, this is done by testing whether additional parameters provide meaningful explanatory gain, with the resulting statistic or p-value summarising the evidence. This motivates defining dissimilarity using frequentist test quantities, in particular a likelihood-ratio (LR) measure that captures improvement in likelihood relative to a null model.

More specifically, let $p_{\mathrm{LR}}(\bs{\xi}^\prime)$ denote the p-value associated with the LR statistic computed from the likelihood evaluated at $\widehat{\mathcal{S}}(\bs{\xi}^\prime)$, and define
\begin{equation}\label{eq:LR_dissimilarity}
d_{\mathrm{LR}}\left(\mathcal{S}(\bs{y}), \widehat{\mathcal{S}}(\bs{\xi}^\prime)\right)
= \log_{10} p_{\mathrm{LR}}(\bs{\xi}^\prime)
\end{equation}
where we have defined 
\begin{equation*}
p_{\mathrm{LR}}(\bs{\xi}^\prime) =  \textrm{Pr}\left( \chi^2_{P^\prime} > \Lambda(\bs{\xi}^\prime) \right) \quad \text{with} \quad 
\Lambda(\bs{\xi}^\prime) = -2 \left\{ \ell_0 - \ell(\bs{\xi}^\prime \mid \bs{y}) \right\}
\end{equation*}
Here $\ell_0$ and $\ell(\bs{\xi}^\prime \mid \bs{y})$ denote the log-likelihoods under the null model and the model specified by $\bs{\xi}^\prime$, respectively and $\ell(\bs{\xi}^\prime \mid \bs{y})$ is the log-likelihood eveluated in correspondence of the maximum lkelihood estimated under model $ \bs{\xi}^\prime$. Under standard regularity conditions, $\Lambda(\bs{\xi}^\prime)$ is asymptotically $\chi^2_{P^\prime}$-distributed, with $P^\prime$ the number of active parameters in $\bs{\xi}^\prime$.

The LR-based dissimilarity provides a measure of improvement over the null model. Taking $\log_{10} p_{\mathrm{LR}}(\bs{\xi}^\prime)$ yields a continuous scale of evidence against the null that integrates directly into the similarity-driven proposal \eqref{eq:similarity_driven}. Its defining distinction is that it is expressed directly in terms of the likelihood evaluated at $\widehat{\mathcal{S}}(\bs{\xi}^\prime)$, thereby accounting explicitly for both goodness-of-fit and model complexity in accordance with classical asymptotic theory \citep{riedle2020reconceptualizing}. As p-values are scale-free and adjusted for model dimension, this choice enables meaningful comparisons across candidate models of differing complexity. The logarithmic transformation further stabilizes extreme values of the test statistic, improving numerical robustness.

Computationally, the cost of evaluating $d_{\mathrm{LR}}(\cdot \, ,\cdot)$ is modest in many settings and comparable to residual-based calculations. Robust, heteroscedasticity-consistent, or permutation-based LR tests may be employed when likelihood assumptions are violated, without modifying the structure of the proposal.

When dealing with regression settings, a natural and interpretable choice is based on the classical F-test for nested linear models. Denote by $p_{\mathrm{F}}(\bs{\xi}^\prime)$ the p-value for testing improvement over a null model using the F-statistic computed from $\widehat{\mathcal{S}}(\bs{\xi}^\prime)$. We define
\begin{equation}\label{eq:F_dissimilarity}
d_{\mathrm{F}}\left(\mathcal{S}(\bs{y}), \widehat{\mathcal{S}}(\bs{\xi}^\prime)\right) = \log_{10} p_{\mathrm{F}}(\bs{\xi}^\prime)
\end{equation}
where
\begin{equation*}
p_{\mathrm{F}}(\bs{\xi}^\prime) = \textrm{Pr}\left(F_{P^\prime,\, n-P^\prime} > F(\bs{\xi}^\prime)\right) \quad \text{with} \quad F(\bs{\xi}^\prime) =
\frac{\big(\mathrm{RSS}_0 - \mathrm{RSS}(\bs{\xi}^\prime)\big)/P^\prime}{\mathrm{RSS}(\bs{\xi}^\prime)/(n - P^\prime)}
\end{equation*}
Here, $\mathrm{RSS}(\bs{\xi}^\prime)$ denotes the residual sum of squares associated with $\widehat{\mathcal{S}}(\bs{\xi}^\prime)$, $\mathrm{RSS}_0$ that of the null model, $P^\prime$ the number of active predictors, and $n$ the sample size. As before, the logarithmic transformation ensures interpretability and facilitates numerically stability.

From a computational standpoint, evaluating $d_{\mathrm{F}}(\cdot \, ,\cdot)$ requires only residual sums of squares and degrees of freedom. When the assumptions underlying the classical F-test are questionable, heteroscedasticity-consistent F-tests, rank-based ANOVA statistics or p-values based on permutation \citep{copt2007robust} may be substituted without altering the structure of the proposal mechanism.

Test-based dissimilarities such as $d_{\mathrm{F}}(\cdot \, ,\cdot)$ and $d_{\mathrm{LR}}(\cdot \, ,\cdot)$  provide principled and automatic ways to bias similarity-driven proposals toward models with meaningful improvements in fit, while retaining  interpretability  and computational tractability.  Alternative distance measures can be used depending on the application, for example simply using a distance between observed and fitted values.

\subsection{Neighborhood Selection}\label{subsec:neigh_sel}

The definition of an appropriate \emph{neighborhood structure} on the discrete state space $\Xi$, tailored to the current model configuration $\bs{\xi}$, is critical for the efficiency, the mixing behavior and computational cost of local MCMC algorithms, including MH \citep{geyer1992practical}. In these algorithms, the base proposal $K(\bs{\xi}^\prime \mid \bs{\xi})$ is typically supported only on a subset of $\Xi$ corresponding to the neighbors of the model defined by $\bs{\xi}$.

Formally, a neighborhood system $N(\bs{\xi})$ assigns to each model, represented by the parameter vector $\bs{\xi} \in \Xi$, a finite set of candidate models $N(\bs{\xi}) \subseteq \Xi$. The corresponding base kernel can be expressed as
\begin{equation*}
K(\bs{\xi}^\prime \mid \bs{\xi}) =
\begin{cases}
\frac{1}{|N(\bs{\xi})|} & \text{if } \bs{\xi}^\prime \in N(\bs{\xi})\\
0 & \text{otherwise}
\end{cases}
\end{equation*}
which defines a \emph{uniform proposal} over the neighborhood $N(\bs{\xi})$, assigning equal probability to each neighboring model configuration. When the neighborhood system is symmetric, i.e., $\bs{\xi}^\prime \in N(\bs{\xi})$ if and only if $\bs{\xi} \in N(\bs{\xi}^\prime)$, the resulting proposal is reversible, which leads to preservation of detailed balance under standard MH updates.

Common choices of $N(\bs{\xi})$ depend on the structure of $\Xi$. In combinatorial or clustering problems, neighborhoods are often tailored to reflect valid local modifications of a partition, such as reassigning a single element to a different cluster, performing split-merge moves, or swapping pairs of elements between clusters. These designs allow the sampler to explore the space of clusterings efficiently while respecting intrinsic combinatorial constraints \citep{jain2004split}. In graphical model selection problems, neighborhoods are often defined via edge additions, deletions or reversals, leading to model-space transitions of varying dimension \citep{green1995reversible}. Finally, for binary inclusion vectors $\Xi = \{0,1\}^P$, a common neighborhood consists of single-component flips, as in Bayesian variable selection schemes based on Gibbs or Metropolis updates \citep{george1993variable}. 

The selection of $N(\bs{\xi})$ involves a trade-off between exploration and exploitation. Small neighborhoods, which allow only minor modifications of the current model, tend to have higher acceptance rates but may hinder global exploration, while larger neighborhoods improve coverage of the state space at the cost of lower acceptance and higher computational burden \citep{gelman1997weak, roberts2001optimal}. Adaptive MCMC strategies can dynamically adjust proposal mechanisms to balance these effects in high-dimensional or complex discrete spaces \citep{andrieu2008tutorial}. 
In the similarity-driven framework of \eqref{eq:similarity_driven}, the neighborhood interacts with the weighting function $g(\bs{\xi}^\prime \mid \bs{\xi})$ to guide proposals toward high-probability regions while preserving MH validity.

\subsection{Adaptive tuning of $\lambda$}

The performance of the proposed similarity-driven sampler critically depends on the tuning parameter $\lambda$, which controls the concentration of the proposal distribution around candidates that yield low discrepancy with the observed data. To automatically optimize the sampler's efficiency, we develop an adaptive scheme that maximizes the acceptance rate during an initial burn-in phase. For a current state $\bs{\xi} \in \Xi$, the proposal probability for moving to a candidate state $\bs{\xi}^\prime$ is set to $Q_{\lambda}(\bs{\xi}^\prime \mid \bs{\xi})$ as in \eqref{eq:similarity_driven}. This formulation encompasses the above presented p-value-based measures from LR-tests or F-tests, as well as alternative metrics such as likelihood-based, rank-based, or distance-based discrepancies. As $\lambda \to 0$, the proposal approaches a uniform distribution over all variables, while large values of $\lambda$ concentrate probability mass on statistically significant variables.

To tune $\lambda$ automatically during sampling, we employ a windowed hill-climbing procedure built on Robbins--Monro stochastic approximation \citep{robbins1951stochastic}. Rather than targeting a pre-specified acceptance rate, the algorithm seeks to increase the empirical acceptance rate as much as possible by adjusting $\lambda$ in log-space, which ensures positivity and numerical stability throughout.
Adaptation proceeds in epochs of $W$ consecutive iterations within the interval $[t_{\text{start}}, t_{\text{end}}] \subset \{1,\ldots,T\}$, where $T$ is the total number of MCMC iterations. At the end of epoch $k$, the empirical acceptance rate is $\alpha^{(k)} = n_{\text{acc}}^{(k)} / W$, where $n_{\text{acc}}^{(k)}$ counts accepted proposals within that epoch. For $k \geq 2$, the log-scale update is
\begin{equation}
\label{eq:rm_update}
\log\lambda^{(k)} = \log\lambda^{(k-1)} + d_k\,\Delta\alpha^{(k)}\,\operatorname{sgn}\!\left(\log\lambda^{(k-1)} - \log\lambda^{(k-2)}\right),
\end{equation}
where $\Delta\alpha^{(k)} = \alpha^{(k)} - \alpha^{(k-1)}$ is the change in acceptance rate between consecutive epochs and $d_k = ck^{-\delta}$ is a diminishing step size with $c > 0$, $\delta \in (0.5, 1]$, satisfying the Robbins--Monro summability conditions $\sum_k d_k = \infty$, $\sum_k d_k^2 < \infty$.

The update \eqref{eq:rm_update} implements a finite-difference hill-climbing strategy on the acceptance rate as a function of $\log\lambda$. The two factors jointly encode a gradient estimate: $\Delta\alpha^{(k)}$ measures whether the acceptance rate improved, and $\operatorname{sgn}(\log\lambda^{(k-1)} - \log\lambda^{(k-2)})$ records the direction of the last step. If the rate improved after moving $\lambda$ in a given direction, the algorithm continues in that direction; if it worsened, the direction is reversed. After each update, $\lambda$ is projected onto $[\lambda_{\min}, \lambda_{\max}]$ to prevent degenerate proposals.

Adaptation is restricted to iterations $t \in [t_{\text{start}}, t_{\text{end}}]$, typically coinciding with the burn-in period. Once $t > t_{\text{end}}$, $\lambda$ is held fixed for all remaining iterations. Freezing the proposal at the end of adaptation ensures that the chain eventually targets the true posterior, as required for ergodicity. The complete procedure is given in Algorithm~\ref{alg:adaptive_lambda}. 
\begin{algorithm}[t]
\caption{Windowed Robbins-Monro Adaptation for $\lambda$}
\label{alg:adaptive_lambda}
\footnotesize
\begin{algorithmic}[1]
\Require Number of MCMC iterations $T$, window size $W$, scale $c$, decay $\delta$, bounds $[\lambda_{\min}, \lambda_{\max}]$, adaptation interval $[t_{\text{start}}, t_{\text{end}}]$
\State \textbf{Initialize:} $\lambda^{(0)}$, $\alpha^{(0)}$, $k \gets 0$, $n_{\text{acc}} \gets 0$, $t_{\text{window}} \gets 0$
\For{$t = 1, 2, \ldots, T$} \Comment{MCMC iterations}
    \State Execute birth-death proposal with parameter $\lambda^{(k)}$
    \State $n_{\text{acc}} \gets n_{\text{acc}} + \mathbb{1}_{\{\text{accept}\}}$
    \State $t_{\text{window}} \gets t_{\text{window}} + 1$
    \If{$t \in [t_{\text{start}}, t_{\text{end}})$ \textbf{and} $t_{\text{window}} = W$}
        \State $k \gets k + 1$
        \State $\alpha^{(k)} \gets n_{\text{acc}} / W$ \Comment{Current window acceptance rate}
        \If{$k > 1$} \Comment{Need two previous rates for update}
            \State $d_k \gets c k^{-\delta}$ \Comment{Step size}
            \State $\Delta \alpha^{(k)} \gets \alpha^{(k)} - \alpha^{(k-1)}$ \Comment{Acceptance change}
            \State $\log(\lambda^{(k)}) \gets \log(\lambda^{(k-1)}) + d_k \, \Delta \alpha^{(k)} \, \text{sgn}\left(\log(\lambda^{(k-1)}) - \log(\lambda^{(k-2)})\right)$ \Comment{Update}
            \State $\log(\lambda^{(k)}) \gets \max\left(\log(\lambda_{\min}), \min\left(\log(\lambda^{(k)}), \log(\lambda_{\max})\right)\right)$ \Comment{Enforce bounds}
            \State $\lambda^{(k)} \gets \exp(\log(\lambda^{(k)}))$
        \EndIf
        \State $n_{\text{acc}} \gets 0$, $t_{\text{window}} \gets 0$ \Comment{Reset window counters}
    \EndIf
\EndFor
\end{algorithmic}
\end{algorithm}

In our implementation, we use a window size of $W=25$ iterations to compute empirical acceptance rates, providing a compromise between responsiveness and variance reduction in the stochastic update. The Robbins-Monro step size is chosen as $d_k = k^{-\delta}$ with $\delta = 0.75$, satisfying the classical conditions for stochastic approximation while ensuring stable yet sufficiently fast adaptation in early iterations. Adaptation is restricted to the burn-in phase and updates are performed in log-space and constrained to $\lambda \in [0.05, 10]$ to guarantee positivity and avoid numerical instability or degenerate proposal behavior. These choices provide a robust and practical default configuration. However, the optimal settings may depend on the dimensionality, correlation structure, and signal strength of the specific application, and can be adjusted accordingly.

Our adaptation scheme belongs to the family of adaptive MCMC methods \citep{andrieu2008tutorial, roberts2009examples} that adjust proposal parameters during burn-in to improve sampler efficiency. While classical results \citep{roberts2007coupling} establish conditions under which adaptation does not compromise asymptotic validity, our approach follows the practical strategy of ceasing adaptation well before the sampling phase begins, ensuring standard MCMC theory applies to the collected samples. The use of acceptance rate as the optimization target is motivated by its direct connection to algorithmic efficiency \citep{gelman1996efficient} and its applicability across diverse proposal mechanisms without requiring knowledge of the target distribution's geometry.

\subsection{Local-Move Proposals}\label{sec:loc_move_prop}

The similarity-driven proposal operates within a pre-specified neighbourhood of the current model, which ensures computational tractability but may miss predictor dependencies such as multicollinearity or hierarchical structure \citep{george1997approaches,hans2007shotgun}, leading to local oscillations and poor exploration. To address this, we introduce a structurally informed, component-specific neighbourhood within the MCMC algorithm that enables coordinated local moves aligned with the underlying dependence structure, improving mixing in a manner consistent with adaptive MCMC approaches that jointly update correlated parameters \citep{peters2010model}; related structured proposals have also proved effective in Bayesian variable selection \citep{liang2022adaptive}, correlated pseudo-marginal methods for GLMs \citep{wan2021adaptive}, and paired-swap moves in high-dimensional model spaces.

Formally, in a regression setting with $P$ candidate predictors, let $\bs{\xi} = (\xi_1, \dots, \xi_P) \in \{0,1\}^P$ denote the inclusion vector, where $\xi_p = 1$ indicates that predictor $p$ is selected. To account for predictor correlations, we define a component-specific neighborhood for the $p$-th variable using a graph $\mathcal{G} = (V, E)$ estimated from the design matrix $\bs{X}$. Nodes correspond to predictors and edges encode conditional dependencies, estimated for example via graphical lasso or neighborhood selection \citep{meinshausen2006variable,friedman2008sparse}. The neighborhood of $\xi_p$ is
\begin{equation*}
N_{\mathcal{G}}(\xi_p) = \{ \xi_q : q \in V \text{ and } (p,q) \in E \}
\end{equation*}
that is, the predictors connected to $p$ in $\mathcal{G}$. This construction allows the sampler to transfer inclusion probability along correlated directions, improving exploration and mixing relative to proposals that ignore dependence.

At each local-move step, we first identify the set of active variables that have at least one inactive neighbor in the graph $\mathcal{G}$, i.e.,
\begin{equation*}
A(\bs{\xi}) = \{ \xi_p : \xi_p = 1 \text{ and } N_{\text{inactive}}(\xi_p) \neq \emptyset \}
\end{equation*}
where $N_{\text{inactive}}(\xi_p) = \{ \xi_q \in N_{\mathcal{G}}(\xi_p) : \xi_q = 0 \}$. This ensures that the proposed swap actually changes the model configuration. If $A(\bs{\xi})$ is empty, no move is made. Otherwise, an active component $\xi_p \in A(\bs{\xi})$ is chosen uniformly, and one inactive neighbor $\xi_q \in N_{\text{inactive}}(\xi_p)$ is selected using the data-informed weights
\begin{equation}\label{eq:local_prop}
w_q(\bs{\xi}^{(p,q)}\mid \bs{\xi}) \propto 
\exp\left\{\Big[-d\left(\mathcal{S}(\bs{y}), \widehat{\mathcal{S}}(\bs{\xi}^{(p,q)})\right)\Big]^{\lambda_{\text{move}}} \right\}
\end{equation}
where $d(\cdot,\cdot)$ measures discrepancy between observations and proposed model, i.e. $\bs{\xi}^{(p,q)} = \bs{\xi} - \bs{e}_p + \bs{e}_q$, with $\bs{e}_p$ and $\bs{e}_q$ unit vectors corresponding to the respective components and $\lambda_{\text{move}}>0$ controls informativeness (see Section~\ref{sec:smlrt_drvn_MH}). Swap-based proposals have similarly been used to enhance exploration in high-dimensional regression \citep{liang2023adaptive}, and more generally improve mixing when predictors are strongly correlated.

This swap preserves the overall model size while redistributing inclusion probability along correlated directions. Hence, the resulting proposal kernel is
\begin{equation*}
Q_{\text{move}}(\bs{\xi}^{(p,q)} \mid \bs{\xi})
= \frac{1}{|A(\bs{\xi})|} 
\frac{w_q(\bs{\xi}^{(p,q)}\mid \bs{\xi})}{\sum_{r : \xi_r \in N_{\text{inactive}}(\xi_p)} w_r(\bs{\xi}^{(p,r)}\mid \bs{\xi})}
\end{equation*}
with the first factor representing the uniform selection of $\xi_p$ and the second factor providing the normalized data-informed probability of choosing $\xi_q$. The reverse transition $Q_{\text{move}}(\bs{\xi} \mid \bs{\xi}^{(p,q)})$ is defined analogously, and the move is accepted according to the MH rule:
\begin{equation*}
\alpha_{\text{move}}(\bs{\xi}, \bs{\xi}^{(p,q)})
= \min \left\{ 1, \frac{\pi(\bs{\xi}^{(p,q)} \mid -) \, Q_{\text{move}}(\bs{\xi} \mid \bs{\xi}^{(p,q)})}{\pi(\bs{\xi} \mid -) \, Q_{\text{move}}(\bs{\xi}^{(p,q)} \mid \bs{\xi})} \right\}
\end{equation*}
Since the swap exchanges one active and one inactive component, the reverse move corresponds to selecting the newly activated component and swapping back, ensuring that the proposal kernel is well defined in both directions.

By incorporating predictor dependencies, this local move improves exploration of correlated regions, enhancing mixing and reducing autocorrelation in high-dimensional settings, and extending locally informed proposals to structured model spaces.

\section{Simulation study}\label{sec:simulations}

We conduct simulations to examine how the informativeness parameter $\lambda$ in the similarity-driven proposal affects the MH acceptance rate in a standard linear regression with response $\bs{y} \in \mathbb{R}^n$ and design matrix $\bs{X} \in \mathbb{R}^{n \times P}$:
\begin{align*}
\bs{y} \mid \bs{\beta}, \sigma^2 &\sim \mathcal{N}(\bs{X} \bs{\beta}, \sigma^2 \bs{I}) \\
\bs{\beta} \mid \sigma^2 &\sim \mathcal{N} \left(\bs{\mu}_0, \sigma^2 \bs{\Lambda}_0^{-1}\right) \\
\sigma^2 &\sim \mathrm{IG}(a_0, b_0)
\end{align*}
corresponding to a Normal-Inverse-Gamma prior on $(\bs{\beta}, \sigma^2)$ with corresponding  hyperparameters $(\bs{\mu}_0, \bs{\Lambda}_0, a_0, b_0)$. Variable selection is performed by introducing a binary inclusion vector $\bs{\xi} = (\xi_1,\ldots,\xi_P)^\top$, where $\xi_p = 1$ indicates inclusion of the $p$-th predictor. Independent Bernoulli priors are assigned as $\xi_p \sim \mathrm{Bernoulli}(\pi)$, $p=1,\ldots,P$, with $\pi \sim \mathrm{Beta}(a_\pi,b_\pi)$.

Let $N(\bs{\xi})$ denote the single-flip neighborhood of a configuration $\bs{\xi} \in \{0,1\}^P$, i.e.,
\begin{equation*}
N(\bs{\xi}) = \Big\{ \bs{\xi}^\prime \in \{0,1\}^P :
\sum_{p=1}^P \mathbf{1}(\xi_p^\prime \neq \xi_p) = 1 \Big\}
\label{eq:neighborhood}
\end{equation*}
Given the current state $\bs{\xi}$, a proposal $\bs{\xi}^\prime \in N(\bs{\xi})$ is generated according to the similarity-driven kernel in \eqref{eq:similarity_driven}.
Additional details are reported in Appendix~\ref{app:add_det_simul}.

Data are generated from the linear model above with $n = 200$ observations and $P = 500$ predictors, of which 5 are active. The design matrix $\bs{X}$ is drawn from a zero-mean multivariate normal with Toeplitz covariance $\rho^{|i-j|}$, $\rho = 0.9$, and standardized columns augmented with an intercept. The true active set $\bs{\xi}^\star$ is selected at random. Intercept and active predictors coefficients are drawn from $\mathcal{N}(0,1)$, and responses from $\bs{y} \sim \mathcal{N}(\bs{X}_{\bs{\xi}^\star} \bs{\beta}_{\bs{\xi}^\star}, \sigma^2 \bs{I}_n)$ with $\sigma^2 = 1$. This produces a sparse, correlated setting with low signal-to-noise.

We assess the empirical behavior of the F-test-based proposal mechanism which uses the dissimilarity function in~\eqref{eq:F_dissimilarity}. Figure~\ref{fig:balls_F_stat} summarizes acceptance rates for 100 equally spaced values of $\lambda \in [0.01, 1.50]$, with each run consisting of 20{,}000 iterations and the first 10{,}000 iterations discarded as burn-in.  For small values of $\lambda$, the acceptance rate increases gradually as proposals become more informed by the F-test. Around $\lambda \approx 0.70$, the acceptance rate rises more sharply, reflecting stronger concentration on influential variables. However, for $\lambda$ approaching 1.1, acceptance rates plateau and rapidly decline, eventually collapsing to zero.  

Excessively large $\lambda$ exaggerates differences in proposal weights, making the sampler nearly deterministic. This restricts exploration to a few candidates and amplifies numerical fluctuations in $d_\mathrm{F}(\cdot,\cdot)$, destabilizing the MH acceptance behavior. Thus, moderate $\lambda$ improves efficiency, while overly large $\lambda$ undermines both exploration and numerical stability.
\begin{figure}[t]
\centering
\includegraphics[width=0.75\linewidth]{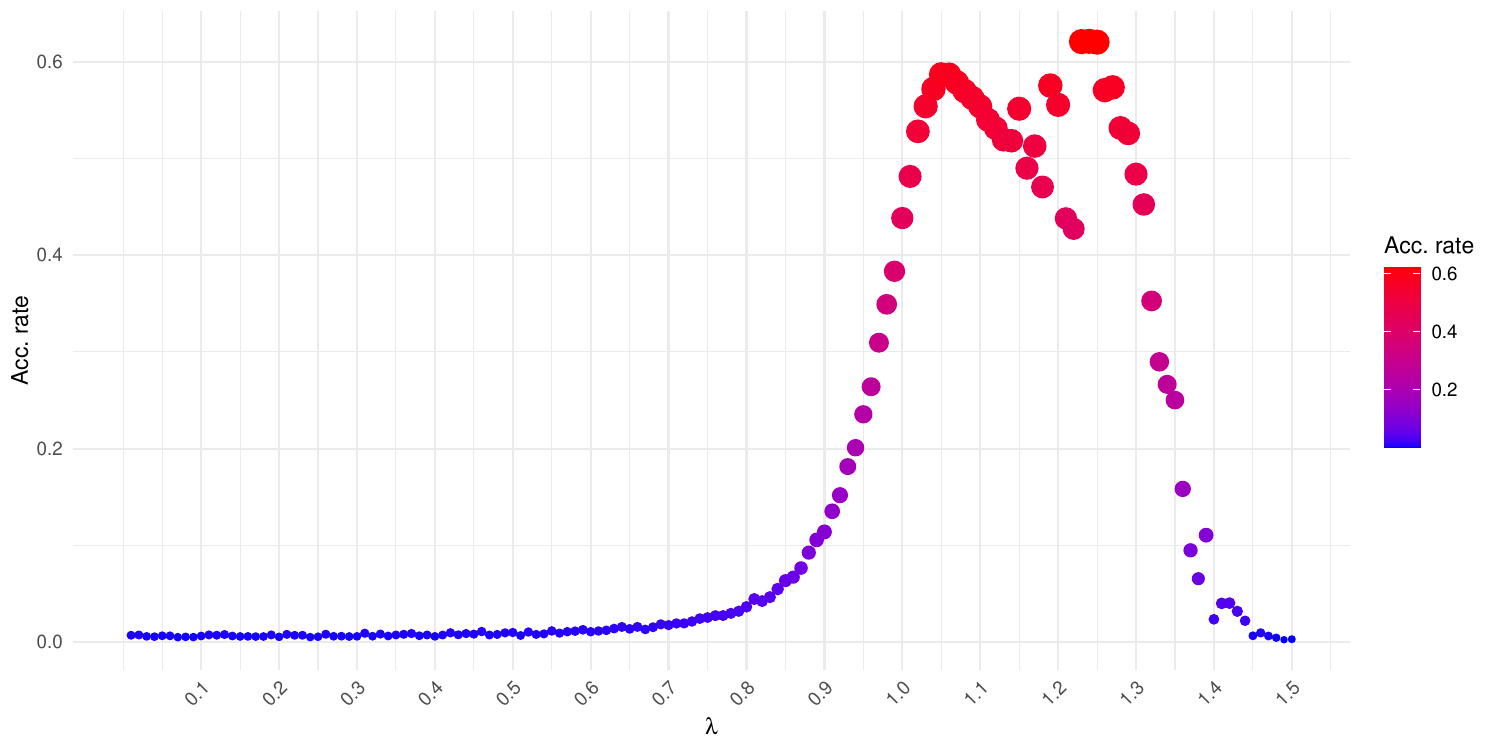}
\caption{Acceptance rates for various choices of $\lambda$ using the F-test proposal.}
\label{fig:balls_F_stat}
\end{figure}

To illustrate the effectiveness of the proposed adaptive tuning scheme, we conduct a controlled experiment on the simulated dataset described above. The goal is not to optimize predictive performance per se, but to verify that the adaptation procedure successfully converges to a value of $\lambda$ that maximizes the acceptance rate.  

In this experiment, we initialize $\lambda$ at a moderate value ($\approx 0.7$) and run the sampler with the windowed Robbins-Monro adaptation (Algorithm~\ref{alg:adaptive_lambda}) to automatically adjust $\lambda$. The empirical acceptance rate is monitored across adaptation windows, and the evolution of $\lambda$ is recorded. The expected behavior is that the algorithm gradually adjusts $\lambda$ toward the value that maximizes the acceptance rate. Figure~\ref{fig:lambda_adaptation} shows the trajectory of $\lambda$ and the corresponding acceptance rates over 100{,}000 iterations (for illustration), with adaptation ceasing at iteration 75{,}000. The figure demonstrates that the adaptation mechanism effectively identifies a near-optimal setting without the need for manual tuning. 
\begin{figure}[t]
\centering
\includegraphics[width=0.49\linewidth]{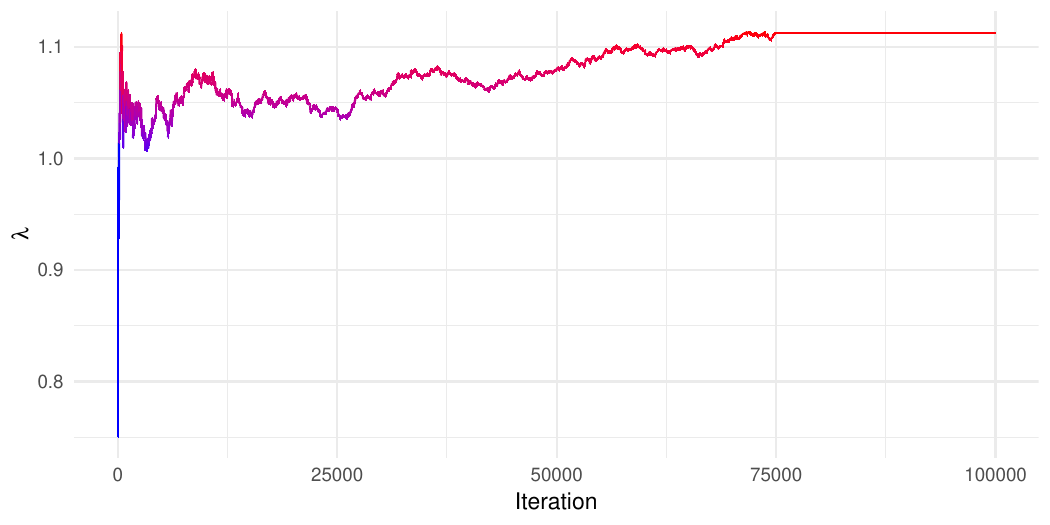}
\includegraphics[width=0.49\linewidth]{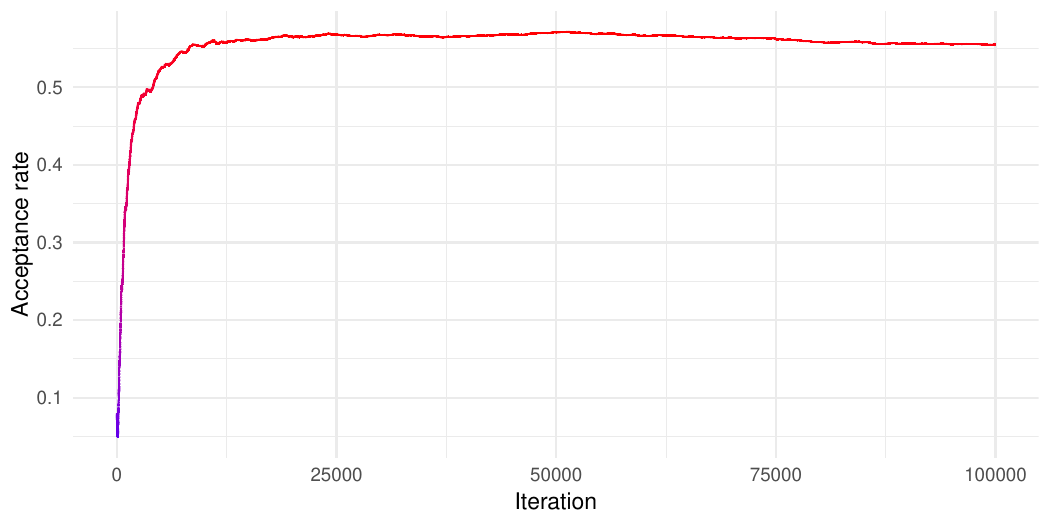}
\caption{Lambda (left) and acceptance rate (right) evolution across iteration with the adaptive scheme.}
\label{fig:lambda_adaptation}
\end{figure}

The F-test-based proposal exhibits robust and interpretable performance over a wide range of $\lambda$ values. By exponentially weighting a transformation of the F-test p-value, the method adaptively emphasizes influential predictors while maintaining adequate stochasticity for effective exploration. However, overly large $\lambda$ values cause a collapse in the effectiveness of the proposal distribution, producing the sharp decline in acceptance rates observed in Figure~\ref{fig:balls_F_stat}. In the next section, we compare these findings with proposals based on LR-test, highlighting similarities and differences in their tail behavior and robustness.

\subsection{Jump distances with local-move proposal}

To account for dependencies among predictors, we introduce an optional local-move step. At each iteration, an active predictor with at least one inactive neighbor, identified via a correlation graph $\mathcal{G}=(V,E)$ estimated from $\bs{X}$ using the graphical lasso \citep{friedman2008sparse}, is selected uniformly, and a swap with one of its inactive neighbors is proposed according to the data-informed dissimilarity weighting in \eqref{eq:local_prop}. This move preserves model size, complements birth-death updates, and facilitates coordinated transitions along correlated directions (see Section~\ref{subsec:neigh_sel}).

To assess the impact of this local-move mechanism on the exploration of the model space, we examine the Hamming distance between successive MCMC states of the inclusion vector $\bs{\xi}$. Under standard single-flip proposals, transitions are restricted to neighborhoods of radius one, whereas the addition of local moves enables coordinated swaps that can induce larger jumps within a single iteration.

We quantify exploration using the Hamming distance between successive inclusion vectors. Distances $d_H = 0,1,2,3$ correspond respectively to no move, a single-flip update, a local swap, or both moves accepted in the same iteration (one `iteration' refers to a composition of a standard MCMC move and a local-swap one). Multi-component transitions are particularly valuable in high-dimensional correlated settings, enabling the sampler to traverse regions that single-variable updates cannot reach.

We run the MCMC algorithm 100 times on the same dataset, each consisting of 20{,}000 iterations with the first 10{,}000 discarded as burn-in. Each run implements the adaptive birth-death proposal as well as the optional local-move step, for which we set $\lambda_{\text{move}} = 1.25$. The local-move neighborhood is typically very small, consisting of only a few highly similar candidate models. In this context, a separate adaptation scheme is unnecessary, as values of $\lambda_{\text{move}}$ close to 1 already provide sufficiently informative weights to guide proposals effectively. At each iteration $l$, we compute the Hamming distance between the current inclusion vector $\bs{\xi}^{(l)}$ and its predecessor $\bs{\xi}^{(l-1)}$:
\begin{equation*}
d_H(\bs{\xi}^{(l)}, \bs{\xi}^{(l-1)}) = \sum_{p=1}^P \mathbf{1}\{\xi_p^{(l)} \neq \xi_p^{(l-1)}\}
\end{equation*}
We record the empirical frequency and proportion of transitions with $d_H \in \{0, 1, 2, 3\}$ across all runs to assess whether the local-move mechanism successfully enables jumps beyond the single-component updates achievable with flip moves alone.

Figure~\ref{fig:hamming_distances} shows similar Hamming distance distributions over 100 runs for both F-test and LR-test proposals. In both cases, local swaps generate frequent multi-component moves ($d_H > 1$), including non-negligible distance-three transitions, indicating effective and coordinated exploration of correlated model spaces. The similarity suggests the mechanism is largely insensitive to the choice of dissimilarity. Additional MCMC diagnostics in Appendix Section~\ref{app:add_det_simul} confirm stable inference and reliable recovery of the data-generating mechanism.
\begin{figure}[t]
\centering
\includegraphics[width=0.49\linewidth]{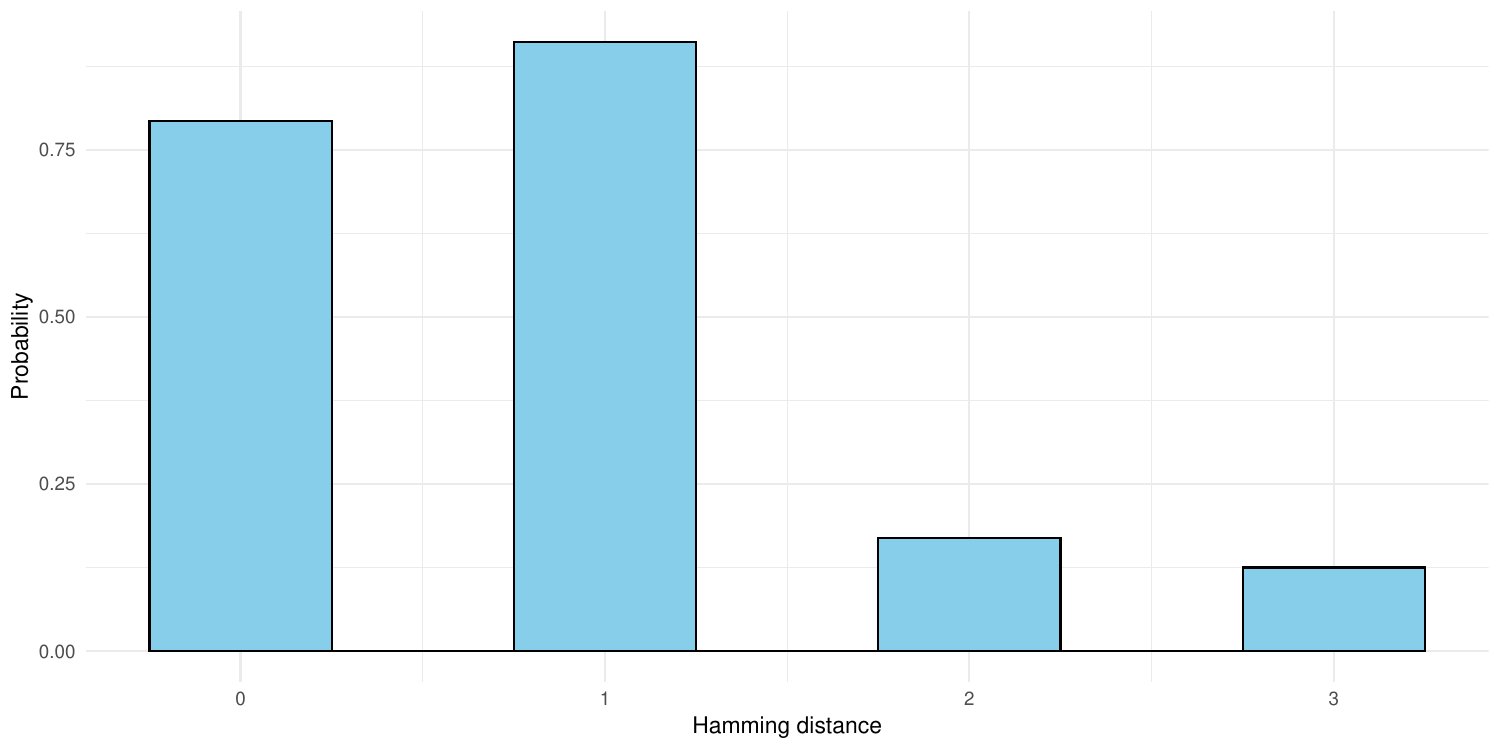}
\includegraphics[width=0.49\linewidth]{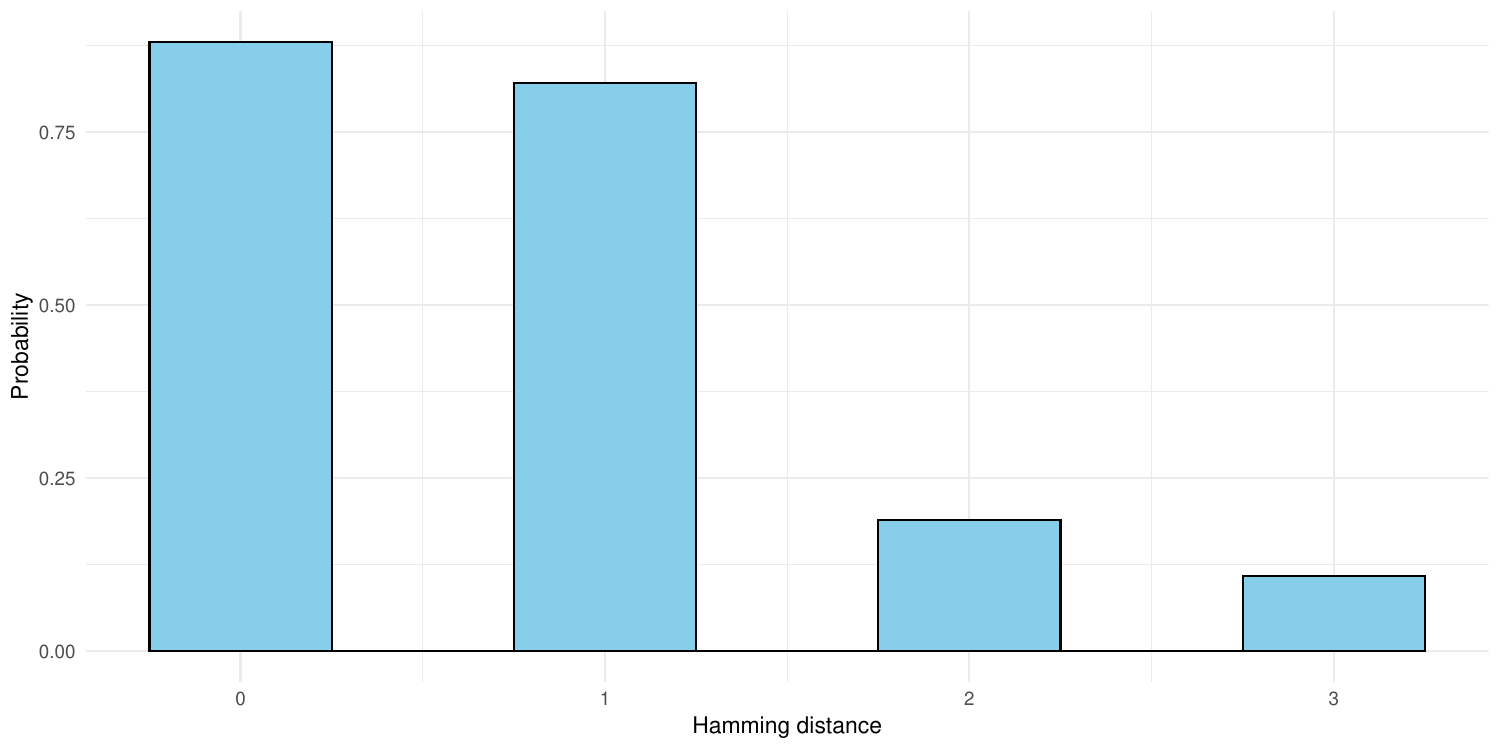}
\caption{Empirical distribution of Hamming distances between consecutive MCMC iterations when including the local-move step across 100 independent runs for the F-test proposal (left) and LR-test proposal (right).}
\label{fig:hamming_distances}
\end{figure}

\section{Dirichlet-Multinomial regression}\label{sec:vs_dm}

Let $\bs{y}_i = (y_{i1},\dots,y_{iJ}) \in \mathbb{N}^J$ denote the vector of counts for $J$ categories corresponding to the $i$-th observation, $i=1,\dots,n$. We model $\bs{y}_i$ using a Multinomial distribution
\begin{equation*}
    \bs{y}_i \mid \bs{\phi}_i \sim \mbox{Multinomial}(y_{i+}, \bs{\phi}_i)
\end{equation*}
where $y_{i+} = \sum_{j=1}^J y_{ij}$, and $\bs{\phi}_i$ is defined on the $(J-1)$-dimensional simplex
\begin{equation*}
    \mathcal{S}^{J-1} = \Big\{ (\phi_{i1},\dots,\phi_{iJ}) : \phi_{ij} \ge 0, \sum_{j=1}^J \phi_{ij} = 1 \Big\}
\end{equation*}
Imposing a conjugate Dirichlet prior on $\bs{\phi}_i$,
\begin{equation*}
    \bs{\phi}_i \mid \bs{\gamma}_i \sim \mbox{Dirichlet}(\bs{\gamma}_i), \quad 
    \bs{\gamma}_i = (\gamma_{i1},\dots,\gamma_{iJ}) \in \mathbb{R}_+^J
\end{equation*}
allows us to integrate out $\bs{\phi}_i$, yielding the Dirichlet-Multinomial (DM) model $\bs{y}_i \mid \bs{\gamma}_i \sim \mbox{DirMult}(y_{i+},\bs{\gamma}_i)$ with
\begin{equation*}
    p(\bs{y}_i \mid \bs{\gamma}_i) = \frac{\Gamma(y_{i+}+1)\Gamma(\gamma_{i+})}{\Gamma(y_{i+} + \gamma_{i+})} \prod_{j=1}^J \frac{\Gamma(y_{ij} + \gamma_{ij})}{\Gamma(y_{ij}+1)\Gamma(\gamma_{ij})}
\end{equation*}
where $\gamma_{i+} = \sum_{j=1}^J \gamma_{ij}$. This model provides greater flexibility than the Multinomial one, particularly when data exhibit overdispersion. Next, we incorporate covariates. Let $\bs{X} = [x_{ip}] \in \mathbb{R}^{n\times P}$ denote the covariate matrix with $P$ predictors for $n$ observations. We link the DM parameters to covariates via a log-linear regression framework:
\begin{equation}\label{eq:DM_gamma_regression}
    \log(\gamma_{ij}) = \beta_{0j} + \sum_{p=1}^P x_{ip} \beta_{pj} \quad \text{for } \quad i=1,\dots,n,\ j=1,\dots,J
\end{equation}
Here, $\beta_{0j}$ is the log-baseline parameter for category $j$, and $\beta_{pj}$ captures the effect of the $p$-th covariate. We assign the following priors to the intercepts:
\begin{equation*}
    \beta_{0j} \mid s_j^2 \sim \mathcal{N}(0, s_j^2)
\end{equation*}
A large value of $s_j^2$ represents a diffuse prior, reflecting weak prior knowledge. In practice, results are robust to this choice; a common default for standardized covariates is $s_j^2 = 10$.

To perform variable selection for each category, we introduce binary inclusion vectors 
$\bs{\xi}_j = (\xi_{1j}, \dots, \xi_{Pj})$, where
\begin{equation*}
    \xi_{pj} = 
    \begin{cases}
        1, & \text{if the $p$-th covariate is included in the model for category $j$,} \\
        0, & \text{otherwise.}
    \end{cases}
\end{equation*}
Conditional on $\bs{\xi}_j$, we assign independent univariate Gaussian priors to the active regression coefficients, i.e., for $p \in \mathcal{P}_j$:
\begin{equation*}
    \beta_{pj} \mid \bs{\xi}_j, r_j^2 \;\stackrel{\text{ind}}{\sim}\; \mathcal{N}(0, r_j^2)
\end{equation*}
where $r_j^2$ is typically chosen large to favor selection of covariates with substantial effects and $\mathcal{P}_j = \{p: \xi_{pj} = 1\}$ indexes the active covariates. The number of coefficients with a prior, $\widetilde{P}_j = \left| \mathcal{P}_j \right|$, depends explicitly on the current inclusion vector $\bs{\xi}_j$, and the model dimension adapts accordingly. Since we need trans-dimensional moves, for posterior computation, we adopt a reversible jump MCMC \citep[RJMCMC;][]{green1995reversible} algorithm; details are given in Section~\ref{subsec:DM_algorithm}. Under this formulation, the linear predictor in Equation~\eqref{eq:DM_gamma_regression} can equivalently be written as
\begin{equation*}
    \log(\gamma_{ij}) = \beta_{0j} + \sum_{p \in \mathcal{P}_j} x_{ip} \beta_{pj}
\end{equation*}
Finally, we assign independent Bernoulli priors to the inclusion indicators:
\begin{equation*}
    \xi_{pj} \mid \pi_{pj} \;\stackrel{\text{ind}}{\sim}\; \text{Bernoulli}(\pi_{pj})
\end{equation*}
with the inclusion probabilities $\pi_{pj}$ themselves modeled hierarchically as $\pi_{pj} \mid a,b \;\stackrel{\text{iid}}{\sim}\; \text{Beta}(a,b)$. Integrating out $\pi_{pj}$ yields a Beta-Binomial prior for each inclusion indicator:
\begin{equation*}
    p(\xi_{pj} \mid a,b) = \frac{\mathrm{Beta}(\xi_{pj}+a, 1-\xi_{pj}+b)}{\mathrm{Beta}(a,b)}
\end{equation*}
so that the prior mean $\mathbb{E}[\xi_{pj}] = a/(a+b)$ controls the baseline probability of including covariates in the model.

Full details on how the LR-based proposal is constructed for this model are available in Appendix Section~\ref{app:add_det_DM}.

\subsection{Reversible Jump MCMC Algorithm}\label{subsec:DM_algorithm}

We adapt, for our purposes, the RJMCMC algorithm proposed by \cite{gagnon2021informed} to jointly sample the posterior distribution of the model parameters $(\bs{\beta}_{0}, \bs{\beta}, \bs{\xi})$, which includes the global intercepts $\bs{\beta}_0$, the category-specific regression coefficients $\bs{\beta}_j$, and the corresponding inclusion indicator vectors $\bs{\xi}_j$. The algorithm alternates between updates of the global intercepts and sequential, category-specific updates of $(\bs{\beta}_j, \bs{\xi}_j)$ for $j = 1, \dots, J$, as summarized in Algorithm~\ref{alg:RJMCMC}.

\begin{algorithm}[t]
\caption{RJMCMC for joint sampling of $(\bs{\beta}_0, \bs{\beta}, \bs{\xi})$}
\label{alg:RJMCMC}
\footnotesize
\begin{algorithmic}[1]
\State \textbf{Input:} Data $\bs{Y}$, design matrix $\bs{X}$, current state $(\bs{\beta}_0^{(r-1)}, \bs{\beta}^{(r-1)}, \bs{\xi}^{(r-1)})$.
%%%%%%%%%%%%%%%%%%%%%%%%%%%%%
\Statex \textbf{Step 1: Update $\bs{\beta}_0$}
\State Propose $\bs{\beta}_0^\prime \sim \mathcal{N}(\bs{\beta}_0^{(r-1)}, \bs{\Sigma}_{\beta_0}^{(r-1)})$.
\State Accept $\bs{\beta}_0^\prime$ with probability
\begin{equation*}
\min\Bigg\{1, \frac{p(\bs{\beta}_0^\prime \mid \bs{Y})}{p(\bs{\beta}_0^{(r-1)} \mid \bs{Y})} \Bigg\}
\end{equation*}
\State Update $\bs{\Sigma}_{\beta_0}^{(r)}$ to target a desired acceptance rate.
%%%%%%%%%%%%%%%%%%%%%%%%%%%%%
\Statex \textbf{Step 2: Update $(\bs{\beta}_j, \bs{\xi}_j)$ sequentially for $j = 1,\dots,J$}

\For{$j = 1$ \textbf{to} $J$}

    \State Propose a new inclusion vector 
    $\bs{\xi}_j' \sim Q_{\lambda_j}(\cdot \mid \bs{\xi}_j^{(r-1)})$

    \State Compute the penalized MLE $\widehat{\bs{\beta}}_j \in \mathbb{R}^{|\bs{\xi}_j'|}$ for the active set of predictors indicated by $\bs{\xi}_j'$, conditioning on $\bs{\beta}_0^{(r-1)}$ and $\{\bs{\beta}_k^{(r-1)} : k \neq j\}$

    \State Compute the Hessian $\bs{H}_j$ of the penalized log-likelihood at $\widehat{\bs{\beta}}_j$

    \State Propose new coefficients:
    \begin{equation*}
        \bs{\beta}_j' \sim \mathcal{N}(\widehat{\bs{\beta}}_j, -\bs{H}_j^{-1})
    \end{equation*}

    \State Accept $(\bs{\beta}_j', \bs{\xi}_j')$ with probability
    \begin{equation*}
        \alpha = \min\Bigg\{1,\;
        \frac{p(\bs{\beta}_j', \bs{\xi}_j' \mid \bs{Y})}
             {p(\bs{\beta}_j^{(r-1)}, \bs{\xi}_j^{(r-1)} \mid \bs{Y})}
        \frac{q(\bs{\xi}_j', \bs{\xi}_j^{(r-1)})}{q(\bs{\xi}_j^{(r-1)}, \bs{\xi}_j')}
        \frac{\mathcal{N}(\bs{\beta}_j^{(r-1)} \mid \widehat{\bs{\beta}}_j^{\text{curr}}, -(\bs{H}_j^{\text{curr}})^{-1})}
             {\mathcal{N}(\bs{\beta}_j' \mid \widehat{\bs{\beta}}_j, -\bs{H}_j^{-1})}
        \Bigg\}
    \end{equation*}
    where $\widehat{\bs{\beta}}_j^{\text{curr}}$ and $\bs{H}_j^{\text{curr}}$ correspond to the current active set $\bs{\xi}_j^{(r-1)}$.
    \State Adapt $\lambda_j$.

\EndFor
\State \textbf{Output:} Updated state $(\bs{\beta}_0^{(r)}, \bs{\beta}^{(r)}, \bs{\xi}^{(r)})$.
\end{algorithmic}
\end{algorithm}

In the first stage, the global vector $\bs{\beta}_0$ is updated via a Gaussian random-walk proposal, with the covariance matrix adaptively tuned to target a desired acceptance rate. This enables efficient exploration of the global parameter space while maintaining stability.  

In the second stage, each category $j$ is updated sequentially. A new inclusion vector $\bs{\xi}_j$ is proposed, and the corresponding regression coefficients $\bs{\beta}_j$ are sampled from a multivariate Gaussian distribution centered at the penalized likelihood estimate, with covariance determined by the local curvature. The category-sequential update modifies one category at a time, improving mixing and acceptance rates in high dimensions. It accommodates changing dimensions, exploits likelihood-informed proposals, and avoids the need for spike-and-slab priors.

To the best of our knowledge, this is the first application of RJMCMC equipped with similarity-driven proposals to variable selection in Dirichlet-Multinomial regression. 
The approach can be readily generalised to any regression setting in which the marginal likelihood is unavailable. In contrast to standard implementations based on uniform or weakly informed proposals, our approach exploits likelihood-based information to preferentially explore covariate configurations that yield substantial improvements in model fit, a feature that is particularly beneficial in high-dimensional settings.

Computational efficiency is achieved through a combination of design choices that reduce the cost of proposal evaluation and improve mixing, including the use of penalized likelihoods to stabilize local model comparisons, reuse of intermediate optimization results across neighboring configurations, and adaptive tuning of continuous-parameter proposals during burn-in. Together, these elements allow the proposed sampler to scale to high-dimensional model spaces while preserving the theoretical validity of the MH framework.

\section{Real data analysis}\label{sec:rda}

We illustrate the proposed methodology using the diet-microbiome dataset of \citet{wu2011linking}, previously analyzed by \citet{chen2013variable}. The study includes dietary intake information for 98 healthy individuals and corresponding gut microbiome profiles obtained from 16S rRNA gene sequencing. Following standard preprocessing steps described in the original studies, dietary variables are normalized and standardized, and highly correlated nutrients are grouped, resulting in 118 representative covariates. Microbial abundances are aggregated at the genus level, and the analysis focuses on 30 genera observed in at least 25 subjects, yielding a $98 \times 30$ response matrix $\bs{Y}$ and a $98 \times 118$ covariate matrix $\bs{X}$. Algorithm~\ref{alg:RJMCMC} is run for 20{,}000 iterations, discarding the first 10{,}000 as burn-in. The specification of the hyperparameters is reported in the Appendix Section~\ref{app:add_det_DM}.

The proposed method identifies 13 diet-microbiome associations with posterior inclusion probabilities (PIPs) exceeding a 0.5 threshold, achieving parsimony while capturing biologically meaningful relationships (Figure~\ref{fig:DM_PIPs}). Among these, 11 associations exhibite PIPs above 0.76, corresponding to a Bayesian false discovery rate (FDR) of 0.05. This level of selectivity contrasts sharply with alternative approaches applied to the same dataset: \citet{wadsworth2017integrative} identify 26 associations, \citet{chen2013variable} find 120, the Bayesian Lasso of \citet{taddy2013multinomial} yields 220, and the correlation-based method of \citet{wu2011linking} detects 711 associations. By reducing the number of identified associations by an order of magnitude compared to earlier methods, our approach directly addresses the interpretability challenge inherent in high-dimensional microbiome studies, delivering a parsimonious set of diet-microbiome relationships that facilitates focused biological investigation and hypothesis generation.
\begin{figure}[t]
    \centering
    \includegraphics[width=0.75\linewidth]{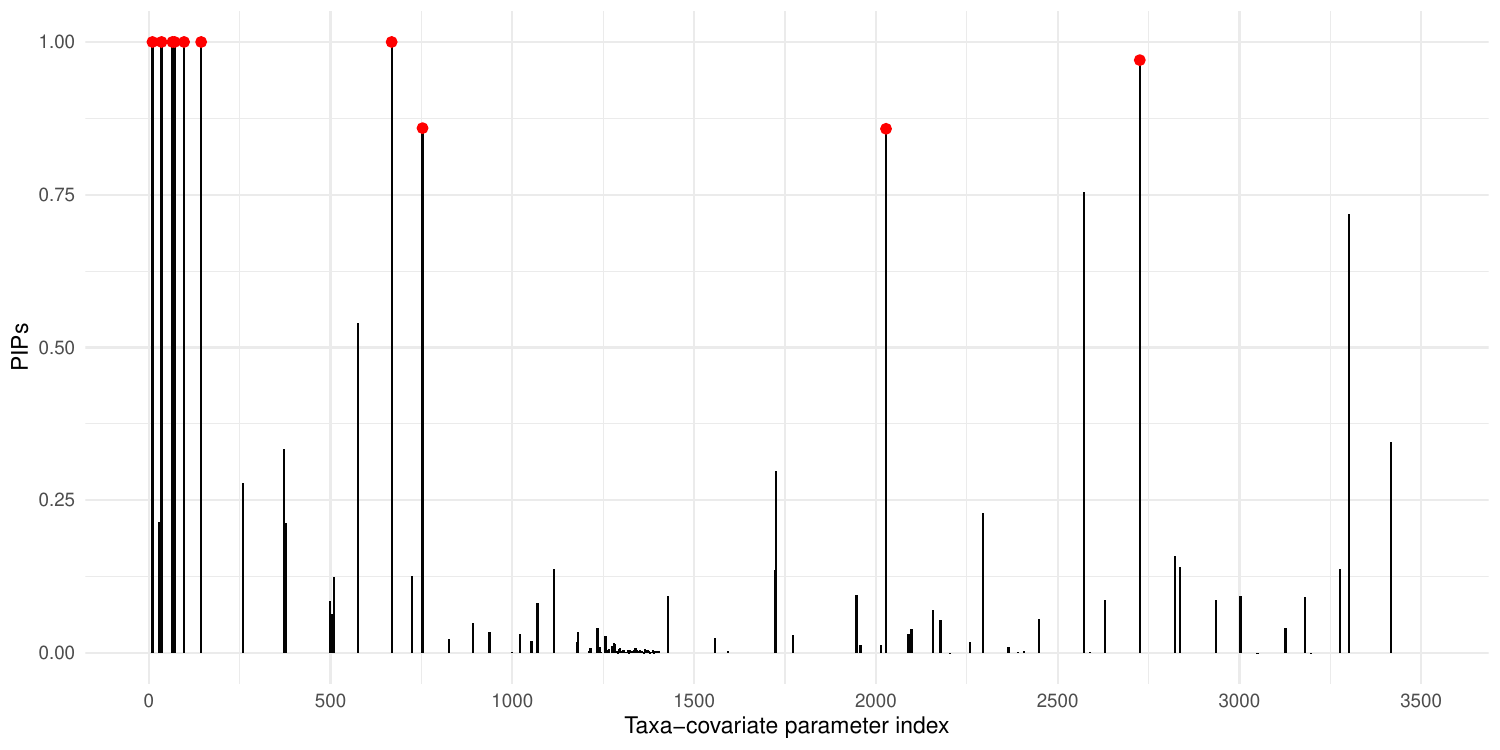}
    \caption{Posterior inclusion probabilities ($\widehat{\xi}_{pj}$). Red dots indicate associations with PIP $> 0.76$, corresponding to Bayesian FDR of 0.05.}
    \label{fig:DM_PIPs}
\end{figure}

Table~\ref{tab:vs_taxa_covariate} summarizes the selected diet-microbiome associations by bacterial order, family, genus, and nutrient. The selected associations are concentrated primarily within the order \textit{Bacteroidales}, with multiple nutrient associations identified for \textit{Bacteroides}, including phosphorus, iodine, vitamin E, food fortification, maltose, and hydroxyproline. Additional associations within \textit{Bacteroidales} involve \textit{Barnesiella} with vitamin B12, \textit{Prevotella} with added germ from wheat, and \textit{Alistipes} with riboflavin B2 excluding vitamin pills. The model also identifies associations within the order \textit{Clostridiales}, linking \textit{Faecalibacterium} to butyric fatty acid and \textit{Phascolarctobacterium} to iodine. Overall, these results illustrate the model's ability to produce a concise and interpretable set of genus-level diet-microbiome associations across distinct bacterial orders. Several selected links involve nutrients or compounds with plausible biological relevance to gut microbial composition and metabolism, while others may represent candidate associations for further biological follow-up. Full biological interpretation and mechanistic context are provided in \citet{wu2011linking,chen2013variable,wadsworth2017integrative}. Finally, more results (including the local move step) are presented in the Appendix, Section~\ref{app:add_det_DM}.
\begin{table}[t]
\centering
\caption{Identified diet-microbiome associations grouped by order}
\footnotesize
\begin{tabular}{lll}
\toprule
\multicolumn{3}{c}{\textit{Order: Bacteroidales}} \\
\midrule
\textbf{Family} & \textbf{Genus} & \textbf{Nutrients} \\
\midrule
Bacteroidaceae     & Bacteroides & Phospherous                  \\
                   &             & Iodine                       \\ 
                   &             & Vitamin E                    \\
                   &             & Food Fortification           \\
                   &             & Maltose                      \\
                   &             & Hydroxyproline               \\                                                      
Porphyromonadaceae & Barnesiella & Vitamin B12                  \\
Prevotellaceae     & Prevotella  & Added Germ from wheats       \\
Rikenellaceae      & Alistipes   & Riboflavin B2 w/o vit. pills \\

\midrule
\multicolumn{3}{c}{\textit{Order: Clostridiales}} \\
\midrule
\textbf{Family} & \textbf{Genus} & \textbf{Nutrients} \\
\midrule
Ruminococcaceae & Faecalibacterium      & Butyric fatty acid \\
Veillonellaceae & Phascolarctobacterium & Iodine             \\
\bottomrule
\end{tabular}%
\label{tab:vs_taxa_covariate}
\end{table}

\section{Discussion}\label{sec:disc}
This paper introduces a flexible and broadly applicable framework for constructing similarity-driven proposals in discrete MCMC settings. By leveraging data-driven discrepancy measures, the approach provides a principled way to bias local moves toward models that better explain the observed data, while retaining exact Metropolis--Hastings validity. This avoids the need for explicit marginal likelihood evaluation, which is often unavailable or computationally prohibitive in complex models.

The proposed methodology bridges a gap between informed MCMC methods in continuous spaces and discrete combinatorial problems. Through the use of similarity-based weighting, the sampler achieves improved exploration of high-probability regions, particularly in high-dimensional and correlated settings such as variable selection. The inclusion of structured local-move proposals further enhances mixing by enabling coordinated updates along dependency structures.

The simulation studies highlight the importance of the tuning parameter $\lambda$ in balancing exploration and exploitation. Moderate values lead to improved acceptance rates and efficient sampling, while overly large values can result in degeneracy and poor mixing. The adaptive Robbins--Monro scheme provides a practical and effective solution for automatically calibrating this parameter, reducing the need for manual tuning.

A key strength of the framework lies in its generality. The proposal mechanism can be adapted to different applications through the choice of discrepancy measure and neighborhood structure, making it suitable for a wide range of discrete inference problems. Moreover, the approach naturally connects to decision-theoretic and generalized Bayesian perspectives, offering a coherent interpretation in terms of loss-based inference.

Future work could explore more sophisticated choices of summary statistics and discrepancy measures, as well as extensions to larger-scale problems and more complex dependency structures. Investigating theoretical properties such as optimal scaling and convergence rates in high-dimensional discrete spaces also remains an important direction. 

\section*{Acknowledgments}
This research was largely conducted while Luca Aiello was a Postdoctoral researcher at the Department of Economics, University of Bergamo, Italy.

\section*{Disclosure statement}
The authors report there are no competing interests to declare.

\bibliographystyle{apalike}
\bibliography{biblio}

\appendix

\section{Additional results on simulation studies}\label{app:add_det_simul}

Since the proposal is generally asymmetric due to the similarity-based weighting, moves are accepted according to the MH acceptance probability, namely
\begin{equation*}
\alpha(\bs{\xi},\bs{\xi}^\prime)
= \min \left\{
1,
\frac{p(\bs{\xi}^\prime \mid \bs{y})}{p(\bs{\xi} \mid \bs{y})} \,
\frac{\exp \left\{\Big[- d\big(\mathcal{S}(\bs{y}), \widehat{\mathcal{S}}(\bs{\xi}^\prime)\big)\Big]^{\lambda}\right\}}
{\exp \left\{\Big[- d\big(\mathcal{S}(\bs{y}), \widehat{\mathcal{S}}(\bs{\xi})\big)\Big]^{\lambda}\right\}} \,
\frac{Z(\bs{\xi})}{Z(\bs{\xi}^\prime)}
\right\}
\label{eq:acceptance_probability}
\end{equation*}

Under the Normal-Inverse-Gamma prior, the marginal likelihood of a model configuration $\bs{\xi}$ admits a closed-form expression,
\begin{equation}
p(\bs{\xi} \mid \bs{y})
= (2\pi)^{-n/2}
\sqrt{\frac{|\bs{\Lambda}_0|}{|\bs{\Lambda}_n|}}
\frac{\Gamma(a_n)}{\Gamma(a_0)}
\frac{b_0^{a_0}}{b_n^{a_n}},
\end{equation}
where the posterior hyperparameters are given by
\begin{align*}
\bs{\Lambda}_n &= \bs{X}_{\bs{\xi}}^\top \bs{X}_{\bs{\xi}} + \bs{\Lambda}_0, \\
\bs{\mu}_n &= \bs{\Lambda}_n^{-1}
\left(\bs{X}_{\bs{\xi}}^\top \bs{X}_{\bs{\xi}}\,\widehat{\bs{\beta}}
+ \bs{\Lambda}_0 \bs{\mu}_0\right), \\
a_n &= a_0 + \frac{n}{2}, \\
b_n &= b_0 + \frac{1}{2}
\left(
\bs{y}^\top \bs{y}
+ \bs{\mu}_0^\top \bs{\Lambda}_0 \bs{\mu}_0
- \bs{\mu}_n^\top \bs{\Lambda}_n \bs{\mu}_n
\right).
\end{align*}

\begin{figure}[htbp!]
    \centering
    \includegraphics[width=0.8\linewidth]{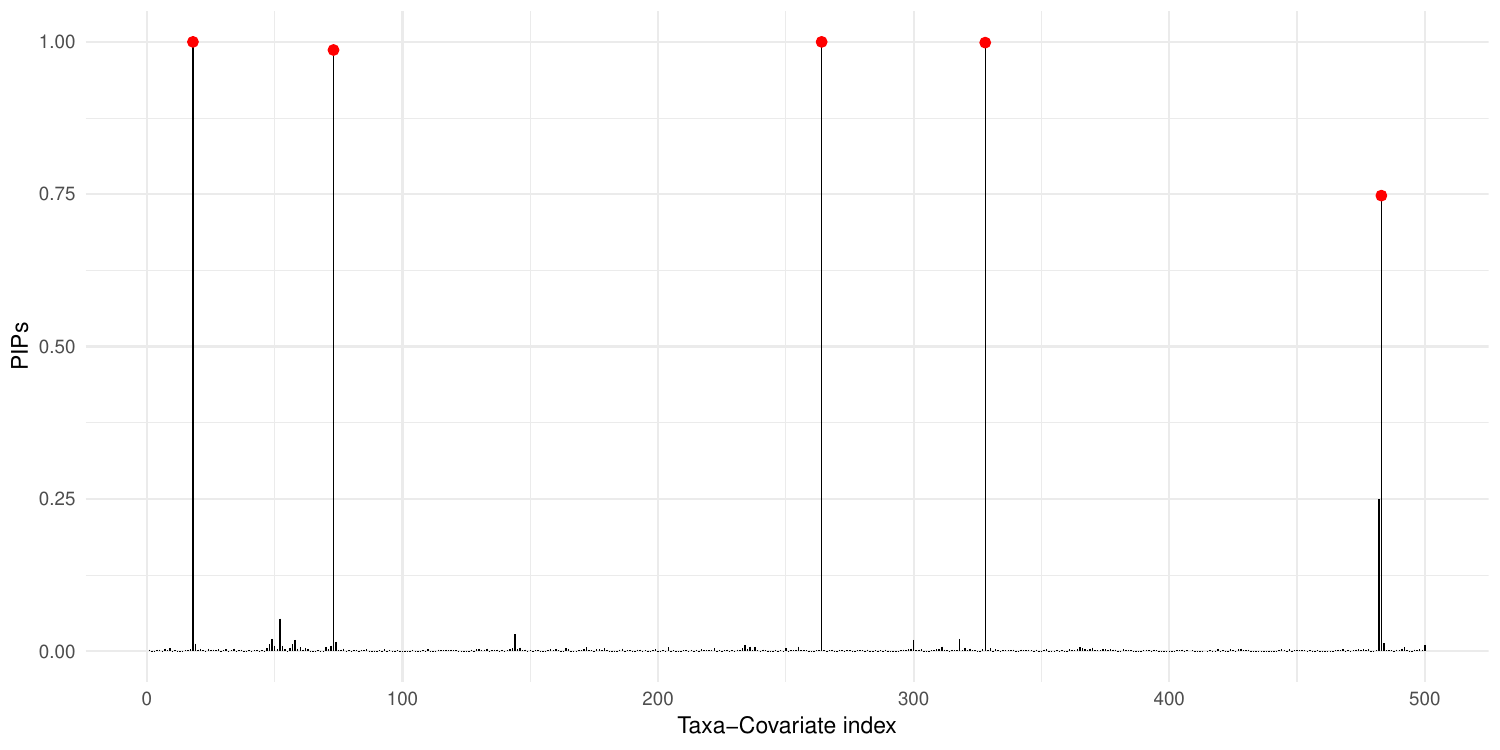}
    \includegraphics[width=0.8\linewidth]{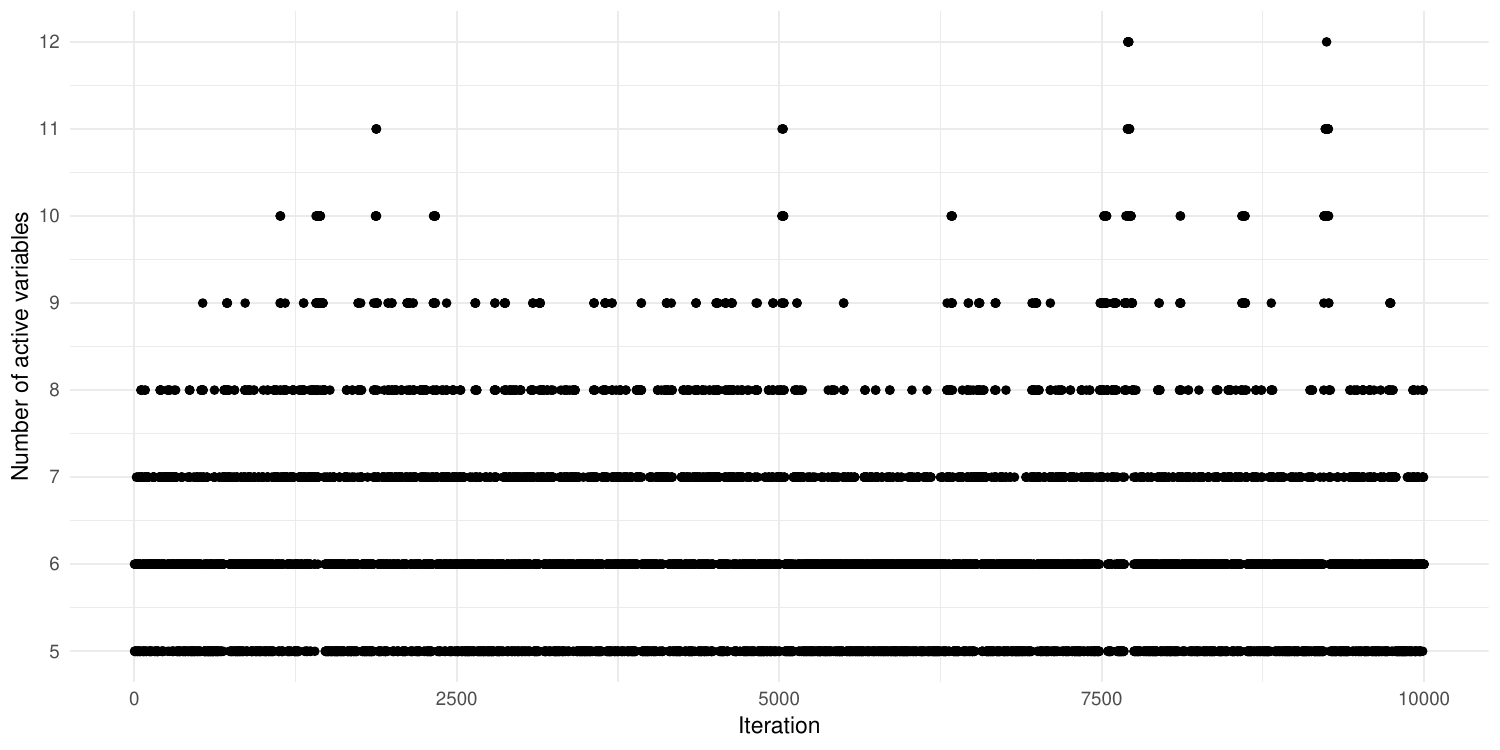}
    \includegraphics[width=0.8\linewidth]{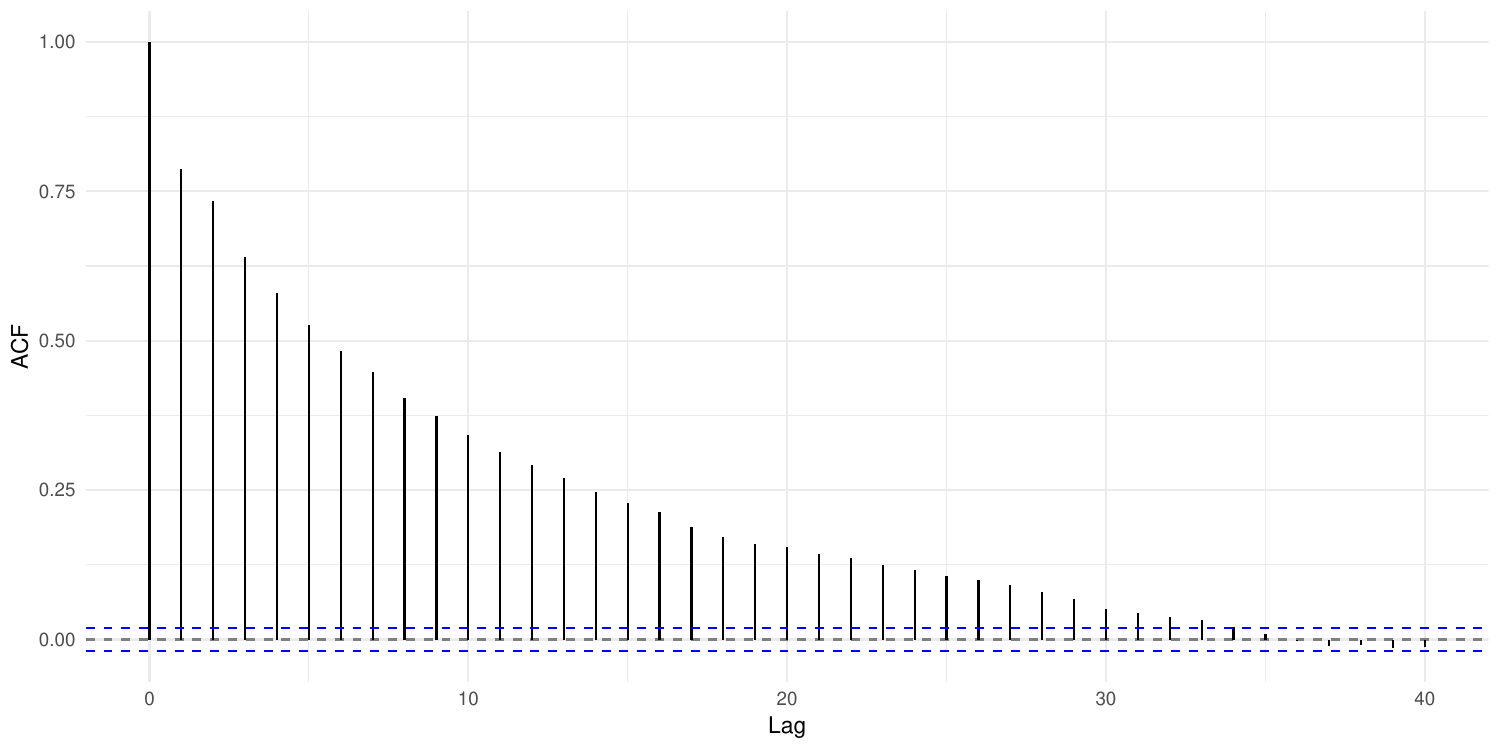}
    \caption{MCMC diagnostics for the adaptive F-stat flip proposal with the additional local-move proposal algorithm: PIPs for all predictors, with truly active variables highlighted in red (top), traceplot of the model size (middle) and autocorrelation function of the model size (bottom).}
    \label{fig:additional_simul_F_stat}
\end{figure}

\begin{figure}[htbp!]
    \centering
    \includegraphics[width=0.8\linewidth]{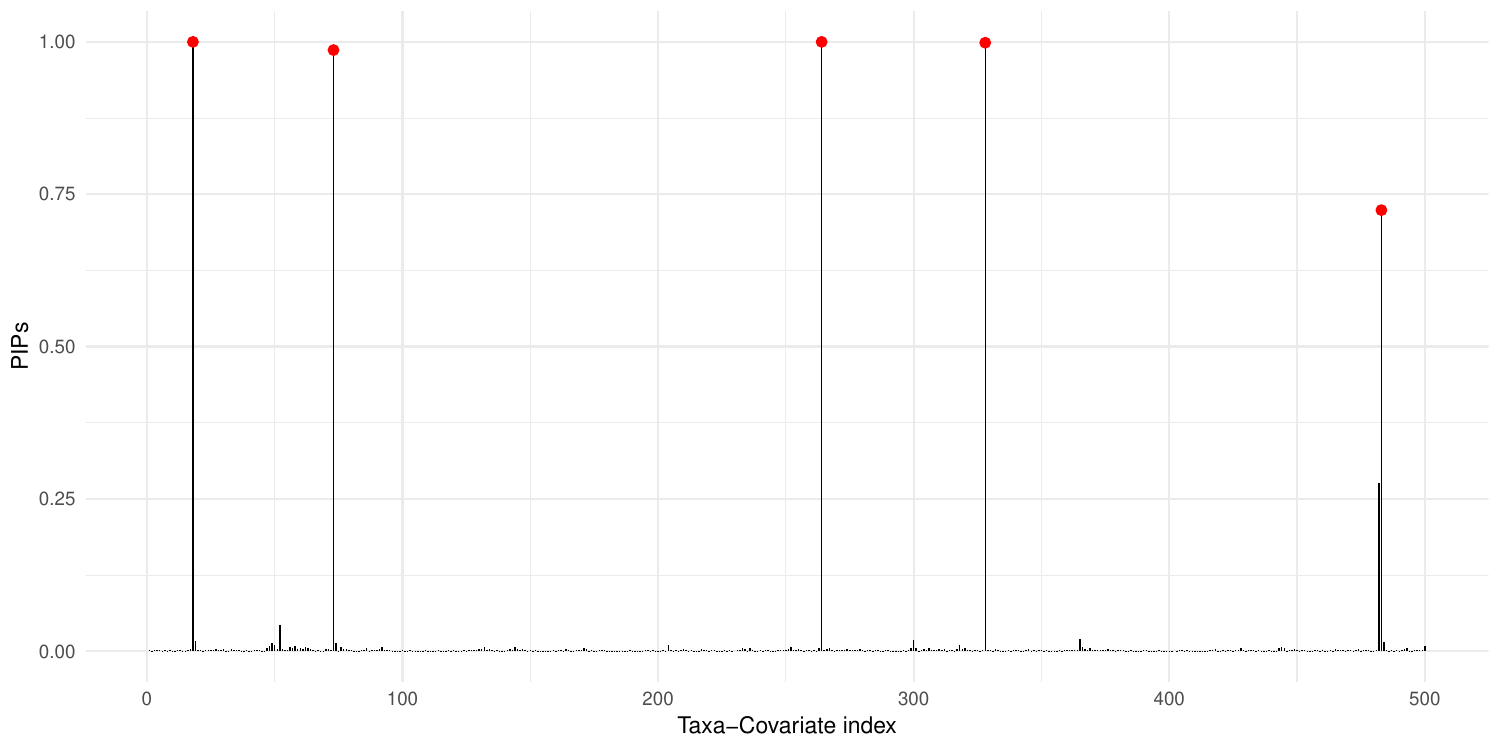}
    \includegraphics[width=0.8\linewidth]{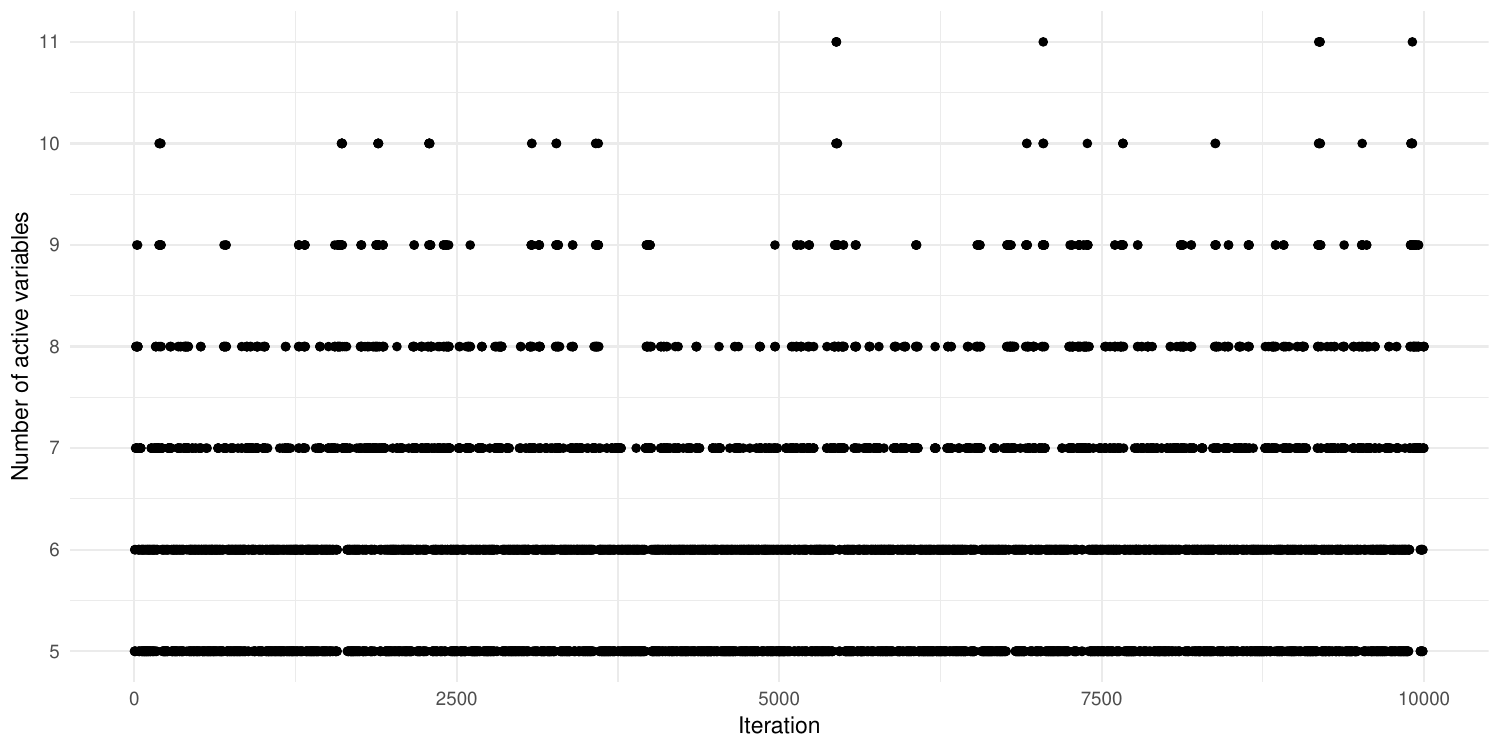}
    \includegraphics[width=0.8\linewidth]{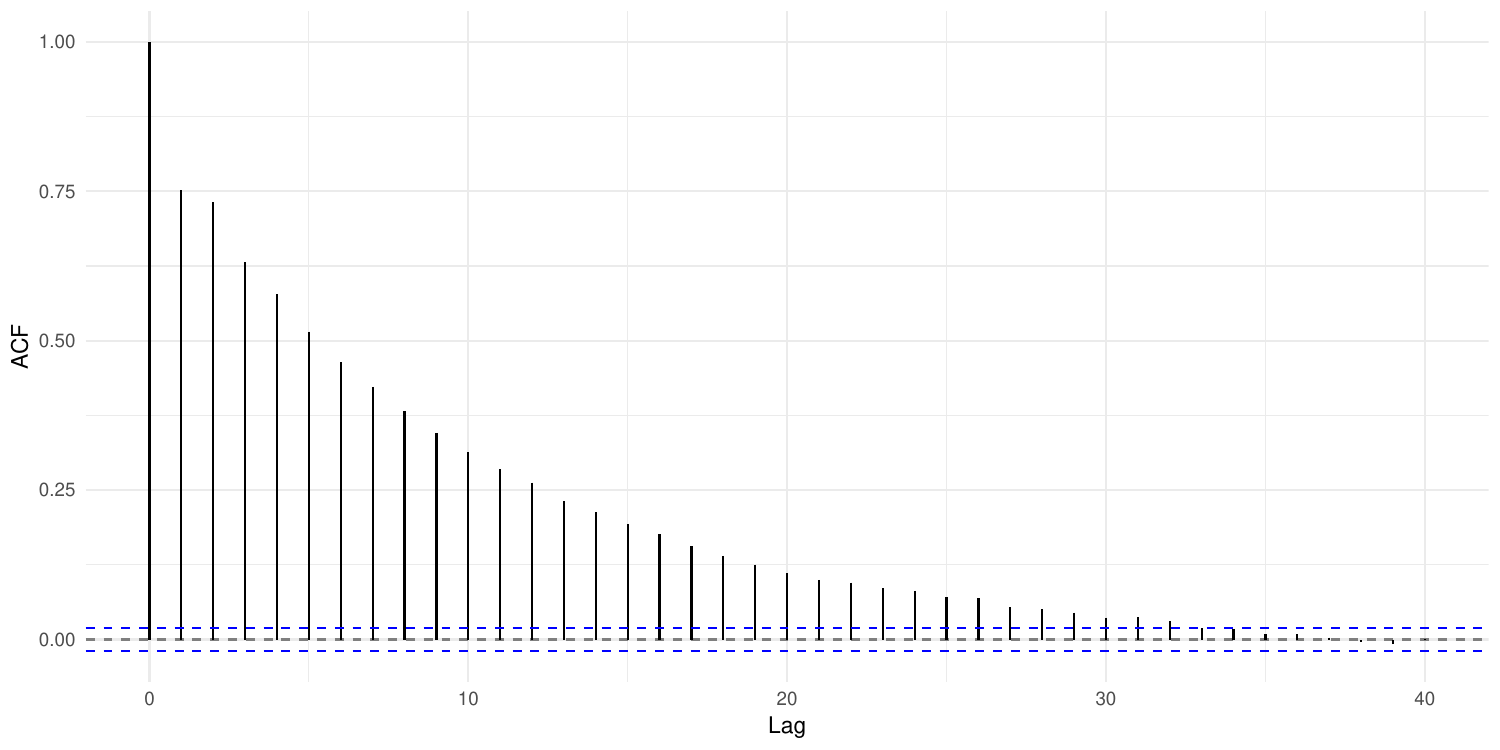}
    \caption{MCMC diagnostics for the adaptive LR flip proposal with the additional local-move proposal algorithm: PIPs for all predictors, with truly active variables highlighted in red (top), traceplot of the model size (middle) and autocorrelation function of the model size (bottom).}
    \label{fig:additional_simul_LR}
\end{figure}

\paragraph{Likelihood Ratio Based Proposal}

Here we examine the empirical behavior of the LR-test-based proposal mechanism, constructed from the dissimilarity measure defined in equation~\eqref{eq:LR_dissimilarity}. Analogously to the F-test approach, this proposal assigns weights to candidate variables according to the evidence provided by the LR-test from the corresponding nested model comparison. In particular, variables that yield greater improvements in model fit, measured through smaller LR-test p-values, receive higher weight. As before, the concentration parameter $\lambda$ modulates the sharpness of this weighting: small values of $\lambda$ yield nearly uniform proposals, whereas larger values concentrate probability mass on variables with stronger evidence.

Figure~\ref{fig:balls_LR} summarizes the acceptance behavior of the LR-test-based proposal across 100 equally spaced values of $\lambda \in [0.01, 1.50]$, using the same MCMC settings as in the F-test-based experiment. The qualitative behavior closely mirrors that observed for the F-test proposal. For small $\lambda$, acceptance rates increase gradually as proposals become more informative, tracking the improved ability of the LR-test statistic to guide model exploration. Around $\lambda \approx 0.70$, we again observe a more pronounced rise in acceptance rates, indicating a beneficial concentration of the proposal distribution around influential predictors. However, as $\lambda$ approaches 1.1-1.2, the acceptance curve levels off before undergoing a sharp and irregular decline, ultimately falling to zero for sufficiently large $\lambda$.

The mechanism behind this collapse is the same as in the F-test case. Large $\lambda$ values exponentially amplify differences in the LR-test-based dissimilarities, $\exp \left\{\big[- d_\mathrm{LR}\big(\mathcal{S}(\bs{y}), \widehat{\mathcal{S}}(\bs{\xi})\big)\big]^{\lambda}\right\}$, forcing the proposal distribution to become nearly deterministic. In this regime, the sampler repeatedly proposes the same (or nearly the same) variable, severely limiting exploration of the model space. As a result, proposals tend to revisit previously explored models or introduce negligible improvement, yielding very low acceptance probabilities. Moreover, the exponential transformation magnifies minor numerical fluctuations in the LR-test, causing unstable and erratic acceptance behavior for large $\lambda$. Thus, as with the F-test proposal, moderate values of $\lambda$ achieve the best trade-off between concentration on meaningful variables and adequate stochasticity, while excessively large $\lambda$ undermine robustness and numerical stability.

\begin{figure}[htbp!]
\centering
\includegraphics[width=0.75\linewidth]{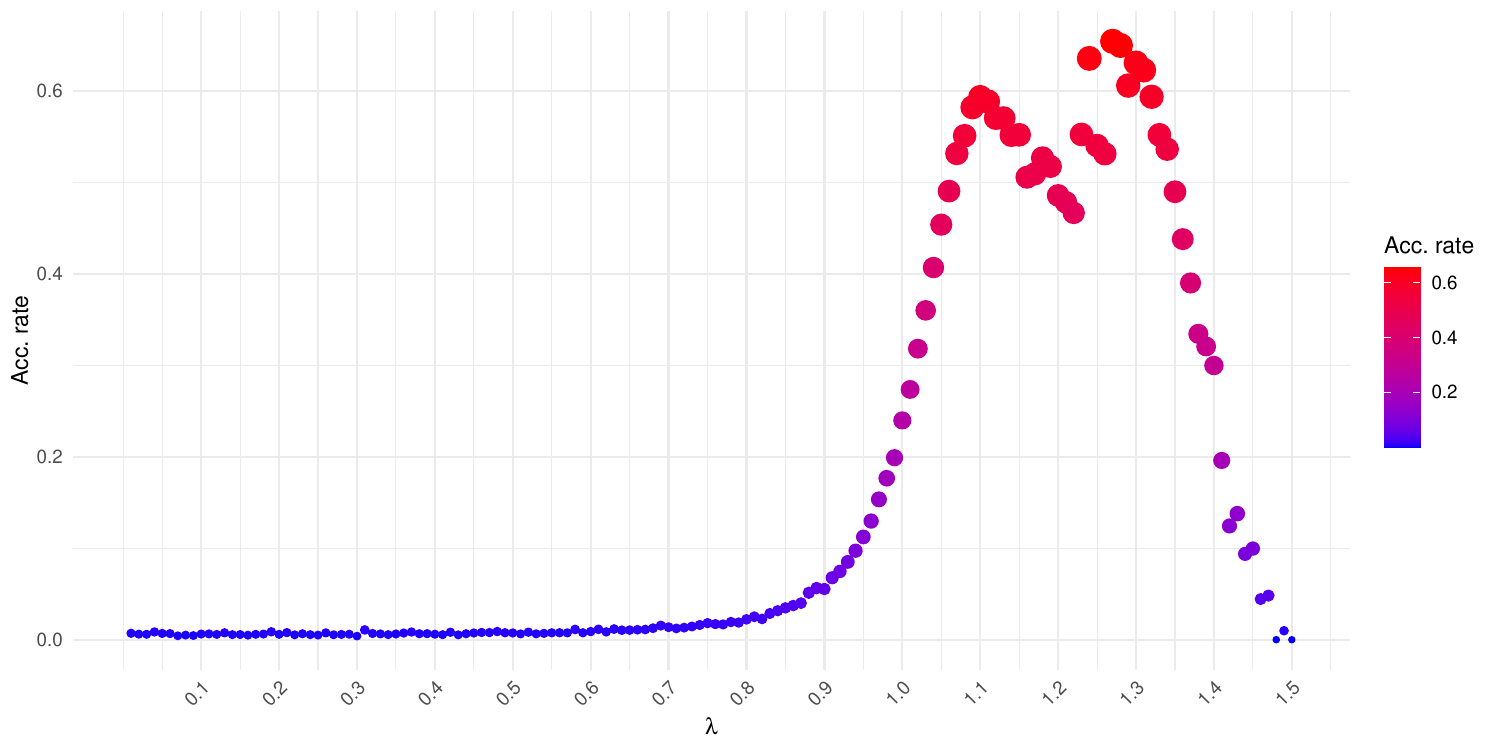}
\caption{Acceptance rates for various combinations of $n$, $P$, and $\lambda$ using the LR-test proposal.}
\label{fig:balls_LR}
\end{figure}

To further evaluate the adaptive tuning strategy introduced in Algorithm~\ref{alg:adaptive_lambda}, we perform an analogous controlled experiment using the LR-based proposal. As in the F-test setting, the aim is not to optimize predictive performance but to verify that the Robbins-Monro adaptation converges toward the value of $\lambda$ that maximizes the acceptance rate. Recall that $\lambda$ regulates the trade-off between exploration (small $\lambda$) and exploitation (large $\lambda$).

We initialize $\lambda$ at a moderate value (approximately 0.7) and apply the windowed Robbins-Monro update over the designated adaptation period. Figure~\ref{fig:lambda_adaptation_LR} presents the evolution of $\lambda$ over 100{,}000 iterations (left) and the corresponding acceptance rates (right), with the adaptation period spanning iterations 100 to 75{,}000. The behavior closely parallels the F-test case: the adaptation mechanism gradually steers $\lambda$ toward a stable, near-optimal region that maximizes acceptance. This confirms that the adaptive scheme effectively identifies suitable values of $\lambda$ even when the underlying proposal is driven by LR-test, eliminating the need for manual tuning.

\begin{figure}[htbp!]
\centering
\includegraphics[width=0.49\linewidth]{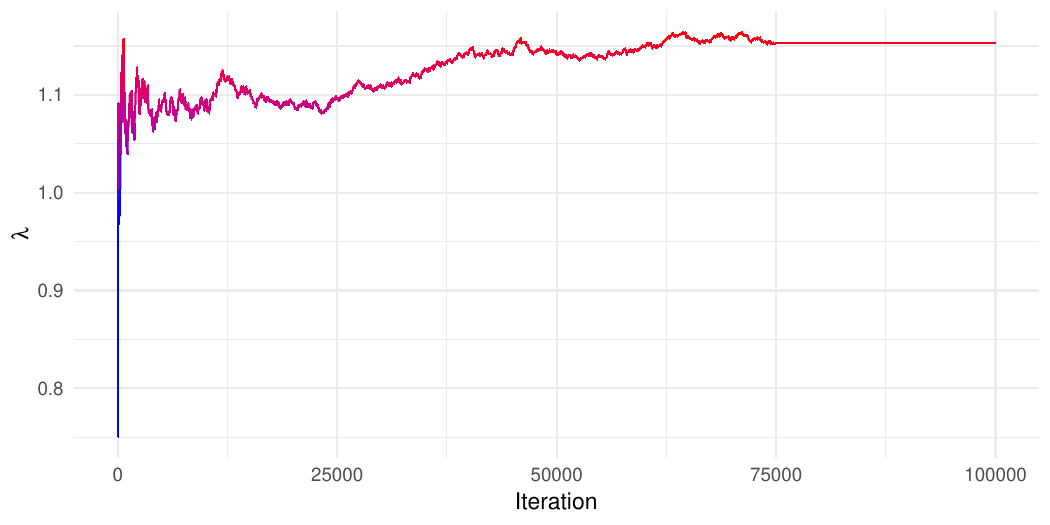}
\includegraphics[width=0.49\linewidth]{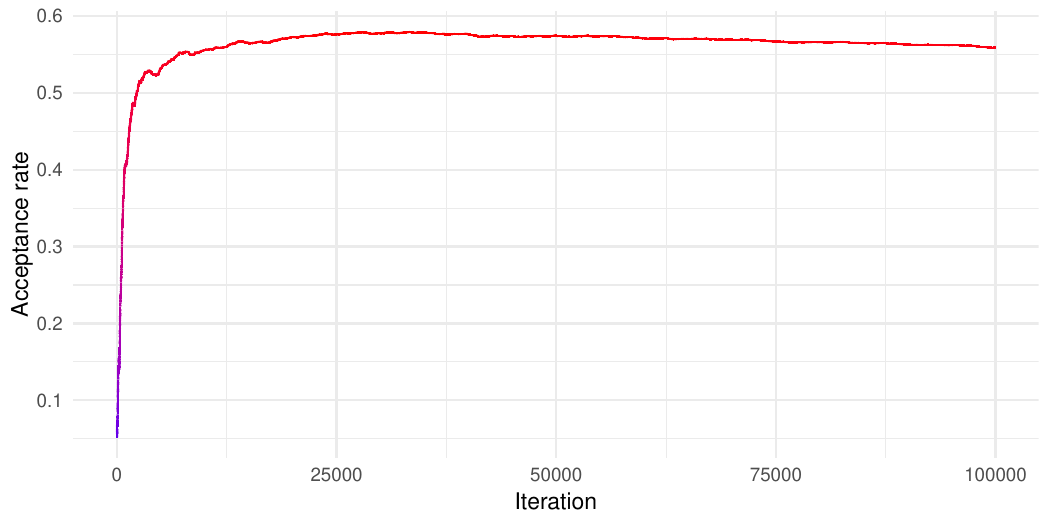}
\caption{Lambda (left) and acceptance rate (right) evolution across iterations for the LR-based adaptive scheme.}
\label{fig:lambda_adaptation_LR}
\end{figure}

In summary, the LR-test-based proposal exhibits behavior that is strongly consistent with the F-test-based mechanism. By exponentially weighting a dissimilarity derived from the LR-test p-value, the method preferentially targets variables that provide substantial improvements in model fit while retaining the stochastic flexibility necessary for efficient exploration. As in the F-test case, however, extreme values of $\lambda$ induce deterministic proposals and exacerbate numerical instabilities, leading to the sharp decline in acceptance rates documented in Figure~\ref{fig:balls_LR}. The adaptive tuning scheme successfully avoids this regime, converging reliably toward values of $\lambda$ that balance exploration and exploitation across different problem instances.

\section{Additional Details on the DM regression}\label{app:add_det_DM}

This appendix collects technical details underlying the construction of the similarity-driven proposal for the Dirichlet-Multinomial (DM) model used in Section~\ref{sec:vs_dm}. In particular, we describe the baseline likelihood employed as a reference for model comparison, the penalized maximum likelihood estimation used to evaluate proposed inclusion configurations, and the resulting update of the Dirichlet concentration parameters.

Under the baseline model with no covariates, the Dirichlet concentration parameters for observation $i$ and category $j$ are denoted by $\gamma_{ij}^{(0)}$, with row sums $\gamma_{i+}^{(0)} = \sum_{j=1}^J \gamma_{ij}^{(0)}$. The associated log-likelihood, used as a global reference for comparing candidate models, is given by
\begin{equation}\label{eq:baseline_loglik}
\ell_0 = \sum_{i=1}^n \Bigg[
\log \Gamma(\gamma_{i+}^{(0)}) - \log \Gamma(y_{i+} + \gamma_{i+}^{(0)})
+ \sum_{j=1}^J \left\{ \log \Gamma(y_{ij} + \gamma_{ij}^{(0)}) - \log \Gamma(\gamma_{ij}^{(0)}) \right\}
\Bigg].
\end{equation}

For a proposed inclusion configuration, regression coefficients are estimated by penalized maximum likelihood. Let $\bs{\xi}_j^\prime$ denote the proposed inclusion vector for category $j$, and let $\bs{\beta}_j$ be the corresponding vector of active regression coefficients. Conditioning on the current values of the intercepts $\bs{\beta}_0^{(r-1)}$ and on the regression coefficients of the remaining categories, the penalized estimator is defined as
\begin{equation*}
\widehat{\bs{\beta}}_j =
\arg\max_{\bs{\beta}_j \in \mathbb{R}^{|\bs{\xi}_j^\prime|}}
\Bigg\{
\ell_j\big(\bs{\beta}_j \,\big|\, \bs{\beta}_0^{(r-1)}, \{\bs{\beta}_k^{(r-1)} : k \neq j\}, \bs{\xi}^\prime\big)
- \frac{c \, |\bs{\xi}_j^\prime|}{2n} \, \bs{\beta}_j^\top \bs{\beta}_j
\Bigg\},
\end{equation*}
where $c>0$ controls the strength of ridge penalization relative to model complexity and $\ell_j(\cdot)$ denotes the log-likelihood contribution associated with category $j$.

Given $\widehat{\bs{\beta}}_j$, the log concentration parameters for category $j$ are updated through the log-linear link
\begin{equation*}
\widehat{\eta}_{ij} = \beta_{0j}^{(r-1)} + \widetilde{\bs{x}}_i^\top \widehat{\bs{\beta}}_j,
\end{equation*}
where $\widetilde{\bs{x}}_i$ denotes the subvector of the $i$-th row of the design matrix corresponding to the predictors active under the proposed configuration. This ensures dimensional consistency between the regression coefficients and the selected covariates. The corresponding Dirichlet concentration parameters are finally obtained as $\widehat{\gamma}_{ij} = \exp(\widehat{\eta}_{ij})$.

These quantities are used to evaluate the likelihood of the proposed model and to construct likelihood-ratio statistics within the similarity-driven proposal mechanism.

We construct a similarity-driven proposal distribution for the inclusion vectors $\bs{\xi}_j$ that systematically explores the model space by leveraging the LR-test. Rather than adopting a uniform proposal $q_{pj} = 1/P$ for $p = 1, \dots, P$ and $j = 1, \dots, J$, our mechanism favors covariate configurations that substantially improve model fit while maintaining computational efficiency through local updates.

Let $\bs{\beta}_0 = (\beta_{01}, \dots, \beta_{0J})^\top \in \mathbb{R}^J$ denote the vector of category-specific intercepts. Under the baseline model with no covariates, the log concentration parameters for observation $i = 1, \dots, n$ and category $j = 1, \dots, J$ are
\begin{equation*}
\eta_{ij}^{(0)} = \beta_{0j}
\end{equation*}
yielding Dirichlet parameters $\gamma_{ij}^{(0)} = \exp(\eta_{ij}^{(0)})$ with row sums $\gamma_{i+}^{(0)} = \sum_{j=1}^J \gamma_{ij}^{(0)}$ and corresponding baseline model likelihood $\ell_0$.

For each category $j = 1, \dots, J$, we explore the model space by flipping a single component of the inclusion vector $\bs{\xi}_j \in \{0,1\}^P$. Specifically, for a given covariate index $p^\prime$, we define the neighboring configuration
\begin{equation*}
\xi_{pj}^\prime =
\begin{cases}
\xi_{pj}, & p \neq p^\prime \\
1 - \xi_{pj}, & p = p^\prime
\end{cases}
\end{equation*}
and, similarly to the linear regression setup presented in Section~\ref{sec:simulations}, explicitly define the neighborhood for each category specific inclusion vector $\bs{\xi}_j$ as
\begin{equation*}
    N(\bs{\xi}_j) = \Big\{\bs{\xi}_j^\prime : \sum_{p=1}^P \mathbf{1}(\xi_{pj}^\prime \neq \xi_{pj}) = 1\Big\}
\end{equation*}
Let $\bs{\xi}^\prime = (\bs{\xi}_1, \dots, \bs{\xi}_{j-1}, \bs{\xi}_j^\prime, \bs{\xi}_{j+1}, \dots, \bs{\xi}_J)$ denote the full inclusion matrix where only the $j$-th column has been updated. We denote by $\mathcal{P}_j^\prime = \{ p : \xi_{pj}^\prime = 1 \}$ the set of selected predictors for category $j$, with cardinality $\widetilde{P}_j^\prime = |\mathcal{P}_j^\prime|$. The corresponding design submatrix is $\bs{X}_{j\bs{\xi}_j^\prime} \in \mathbb{R}^{n \times \widetilde{P}_j^\prime}$.

To evaluate the plausibility of a proposed inclusion configuration for category $j$, we compute a penalized maximum likelihood estimate (PMLE) of the regression coefficients, conditioning on the current intercepts $\bs{\beta}_{0}^{(r-1)}$ and the coefficient vectors of all other categories, $\{\bs{\beta}_k^{(r-1)} : k \neq j\}$. Let $\bs{\beta}_j \in \mathbb{R}^{|\bs{\xi}_j^\prime|}$ denote the vector of coefficients corresponding precisely to the predictors for which $\xi_{jp}^\prime = 1$, that is, the active set of predictors in the proposed configuration. For each proposed configuration, regression coefficients are estimated via penalized maximum likelihood.
The optimization is performed using L-BFGS with warm starts from previously computed estimates, which are stored in a configuration-indexed memory structure to improve computational efficiency.

Given a proposed inclusion configuration, the corresponding regression coefficients are mapped to updated Dirichlet concentration parameters through the log-linear link implied by the model. This update is performed using only the predictors active in the proposed configuration, ensuring dimensional consistency. The resulting concentration parameters are then used to evaluate the likelihood of the proposed model.

We then form the LR-test statistic comparing the proposed model to the baseline:
\begin{equation*}
\mathrm{LR}_{pj} = -2(\ell_0 - \widehat{\ell})
\end{equation*}
where $\ell_0$ is defined in \eqref{eq:baseline_loglik} and $\widehat{\ell}$ is defined analogously with $\widehat{\gamma}_{ij}$ instead of $\gamma_{ij}^{(0)}$. Under standard asymptotic theory, $\mathrm{LR}_{pj}$ approximately follows a $\chi^2_{\widetilde{P}_j^\prime}$ distribution. The corresponding upper-tail probability is
\begin{equation*}
p\text{-value}_{pj} = \Pr\left(\chi^2_{\widetilde{P}_j^\prime} > \mathrm{LR}_{pj}\right)
\end{equation*}
To transform these test statistics into proposal probabilities, we define for $p = 1, \dots, P$
\begin{equation*}
\log q_{pj} \propto \big[-\log_{10}(p\text{-value}_{pj})\big]^\lambda
\end{equation*}
where the tuning parameter $\lambda > 0$ controls the concentration toward more statistically significant covariates. After normalization,
\begin{equation*}
q_{pj} = \frac{\exp(\log q_{pj})}{\sum_{p=1}^P \exp(\log q_{pj})}
\end{equation*}
we obtain a categorical distribution $\bs{q}_j = (q_{1j}, \dots, q_{Pj})$ that defines the probability of proposing a flip for each covariate in category $j$, conditional on all other inclusion vectors $\{\bs{\xi}_k : k \neq j\}$ and the current parameter values.

The full proposal mechanism for category $j$ proceeds as follows: a single covariate index $p^\prime$ is sampled from the categorical distribution with probabilities $\bs{q}_j$, its inclusion status is flipped to obtain $\bs{\xi}_j^\prime$, and all other inclusion vectors $\{\bs{\xi}_k : k \neq j\}$ remain unchanged. The proposal transition kernel is then
\begin{equation*}
q(\bs{\xi}_j^{(r-1)}, \bs{\xi}_j^\prime) = q_{p^\prime j}
\end{equation*}
where $p^\prime$ is the index at which $\xi_{p^\prime j}^\prime \neq \xi_{p^\prime j}^{(r-1)}$. This construction ensures reversibility of the MH step while systematically concentrating computational effort on local moves that are likely to improve model fit. 

To efficiently compute the log-likelihood under the proposed configuration, we exploit the structure of the DM model to perform an incremental update. Let $\gamma_{ij}^{(r-1)} = \exp(\eta_{ij}^{(r-1)})$ denote the current Dirichlet parameters, with row sums $\gamma_{i+}^{(r-1)} = \sum_{j=1}^J \gamma_{ij}^{(r-1)}$. The proposed row sums are
\begin{equation*}
\widehat{\gamma}_{i+} = \gamma_{i+}^{(r-1)} - \gamma_{ij}^{(r-1)} + \widehat{\gamma}_{ij}
\end{equation*}
The log-likelihood under the proposed configuration can then be computed incrementally as
\begin{equation*}\label{eq:incremental_loglik}
\widehat{\ell} = \ell^{(r-1)} + \Delta_1 + \Delta_2
\end{equation*}
where $\ell^{(r-1)}$ is the log-likelihood at the current state, and
\begin{align*}
\Delta_1 &= \sum_{i=1}^n \Big[\log \Gamma(y_{ij} + \widehat{\gamma}_{ij}) - \log \Gamma(\widehat{\gamma}_{ij}) - \log \Gamma(y_{ij} + \gamma_{ij}^{(r-1)}) + \log \Gamma(\gamma_{ij}^{(r-1)}) \Big] \\
\Delta_2 &= \sum_{i=1}^n \Big[\log \Gamma(\widehat{\gamma}_{i+}) - \log \Gamma(\gamma_{i+}^{(r-1)}) - \log \Gamma(y_{i+} + \widehat{\gamma}_{i+}) + \log \Gamma(y_{i+} + \gamma_{i+}^{(r-1)}) \Big]
\end{align*}
This incremental computation avoids recomputing the entire log-likelihood and scales linearly with the sample size.

In the end here we specify the hyperparameters values used for the data analysis. The penalization constant is set to $c = 1$, with prior hyperparameters $a = 1$ and $b = 9$ for $\xi_{pj}$, yielding a mean prior inclusion probability of 0.1. Prior variances for intercepts $\beta_{0j}$ and regression coefficients $\beta_{pj}$ are set to $r_j^2 = s_j^2 = 10$ across all categories, and the concentration parameter was $\lambda = 1$.

A schematic representation of the model is provided in Figure~\ref{fig:DM_diagram}.
\begin{figure}[htbp!]
    \centering
    \includegraphics[width=\linewidth]{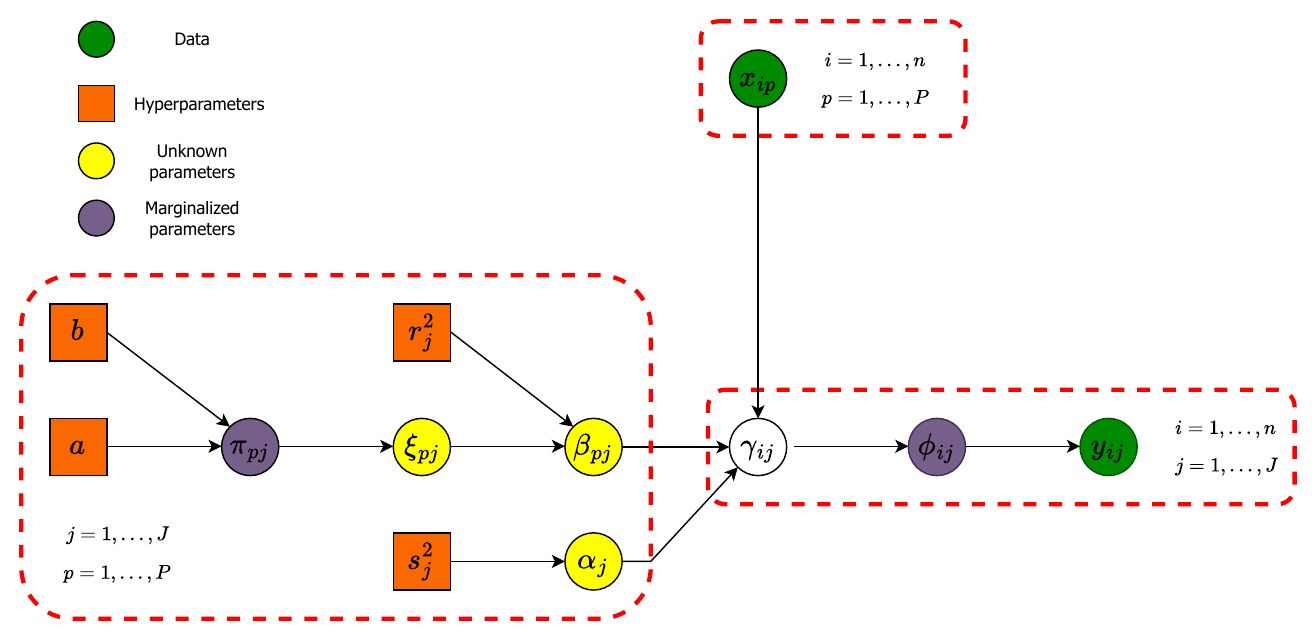}
    \caption{Diagram of the model.}
    \label{fig:DM_diagram}
\end{figure}

\subsection{Additional analysis with local move step}\label{app:DM_local_move}

As an additional analysis, we considered a variant of the RJMCMC algorithm in which the adaptive proposal is augmented with the local move step described in Section~\ref{sec:loc_move_prop} for 10{,}000 iterations, discarding the first 5{,}000 as burn-in.. The purpose of this comparison is to assess whether the strongest diet--microbiome associations identified in the main analysis are robust to a modification of the proposal mechanism. The same preprocessing, model specification, and posterior inclusion probability (PIP) thresholding strategy used in the main real-data analysis were retained.

Figure~\ref{fig:DM_PIPs_move} reports the PIPs obtained under the adaptive proposal with local move step. Compared with the main analysis, this variant yields a more conservative selected model, in the sense that fewer associations exceed the final selection threshold. At the same time, the PIP profile suggests a broader exploration of the model space: posterior mass is assigned to a larger set of candidate taxa--covariate associations, including several associations with non-negligible but sub-threshold PIPs. This indicates that the local move step facilitates movement across nearby models, rather than concentrating posterior exploration only around the strongest selected associations.

In particular, three associations are retained: \textit{Bacteroides} with iodine, \textit{Prevotella} with Added Germ from wheats, and \textit{Phascolarctobacterium} with iodine; see Table~\ref{tab:vs_taxa_covariate_move}. These three associations are all contained in the set selected by the main adaptive proposal, suggesting that they represent the most stable signals across the two implementations. Thus, the local move step leads to a stricter final selection while providing evidence of wider exploration of the posterior model space.
\begin{table}[htbp!]
\centering
\caption{Diet--microbiome associations identified by the adaptive proposal with local move step}
\begin{tabular}{lll}
\toprule
\multicolumn{3}{c}{\textit{Order: Bacteroidales}} \\
\midrule
\textbf{Family} & \textbf{Genus} & \textbf{Nutrients} \\
\midrule
Bacteroidaceae     & Bacteroides & Iodine                       \\ 
Prevotellaceae     & Prevotella  & Added Germ from wheats       \\
\midrule
\multicolumn{3}{c}{\textit{Order: Clostridiales}} \\
\midrule
\textbf{Family} & \textbf{Genus} & \textbf{Nutrients} \\
\midrule
Veillonellaceae & Phascolarctobacterium & Iodine             \\
\bottomrule
\end{tabular}%
\label{tab:vs_taxa_covariate_move}
\end{table}

\begin{figure}[htbp!]
    \centering
    \includegraphics[width=0.75\linewidth]{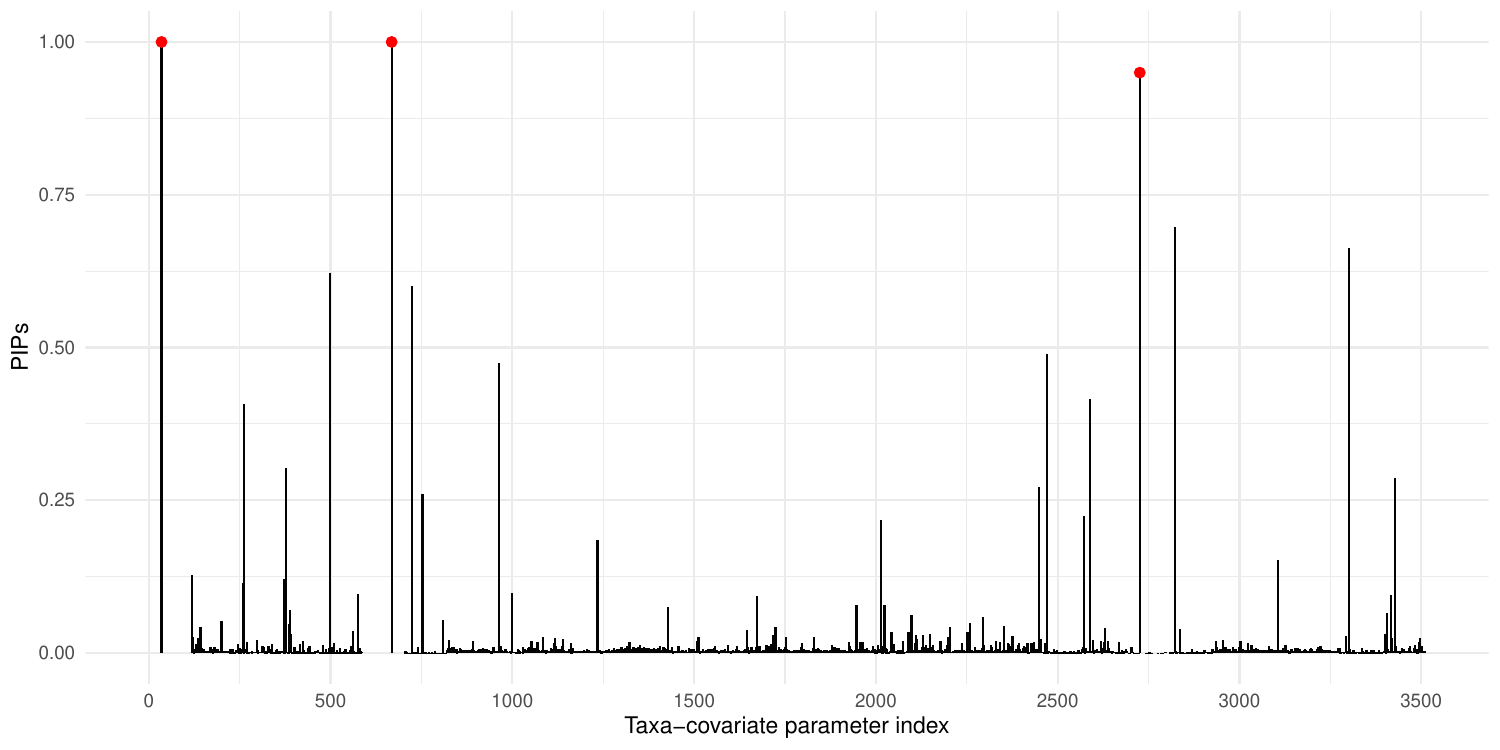}
    \caption{Posterior inclusion probabilities ($\widehat{\xi}_{pj}$) obtained with the adaptive proposal augmented by the local move step. Red dots indicate selected associations under the same decision rule used in the main real-data analysis.}
    \label{fig:DM_PIPs_move}
\end{figure}

\subsection{Diagnostics details of the MCMC}

Figure~\ref{fig:DM_traceplot_autocorrelation} shows rapid convergence of the RJMCMC algorithm when only the $\lambda$ adaptation is included. The trace plot for the number of active associations exhibits stable mixing, and the autocorrelation function decays quickly, indicating efficient exploration of the posterior distribution. Notably, convergence was achieved within 20{,}000 iterations without thinning, compared to the 500{,}000 iterations with 100-fold thinning required by \citet{wadsworth2017integrative}.  The efficiency gains are particularly valuable for microbiome applications, where permutation-based inference or cross-validation procedures often require multiple algorithm runs.
\begin{figure}[htbp!]
    \centering
    \includegraphics[width=0.495\linewidth]{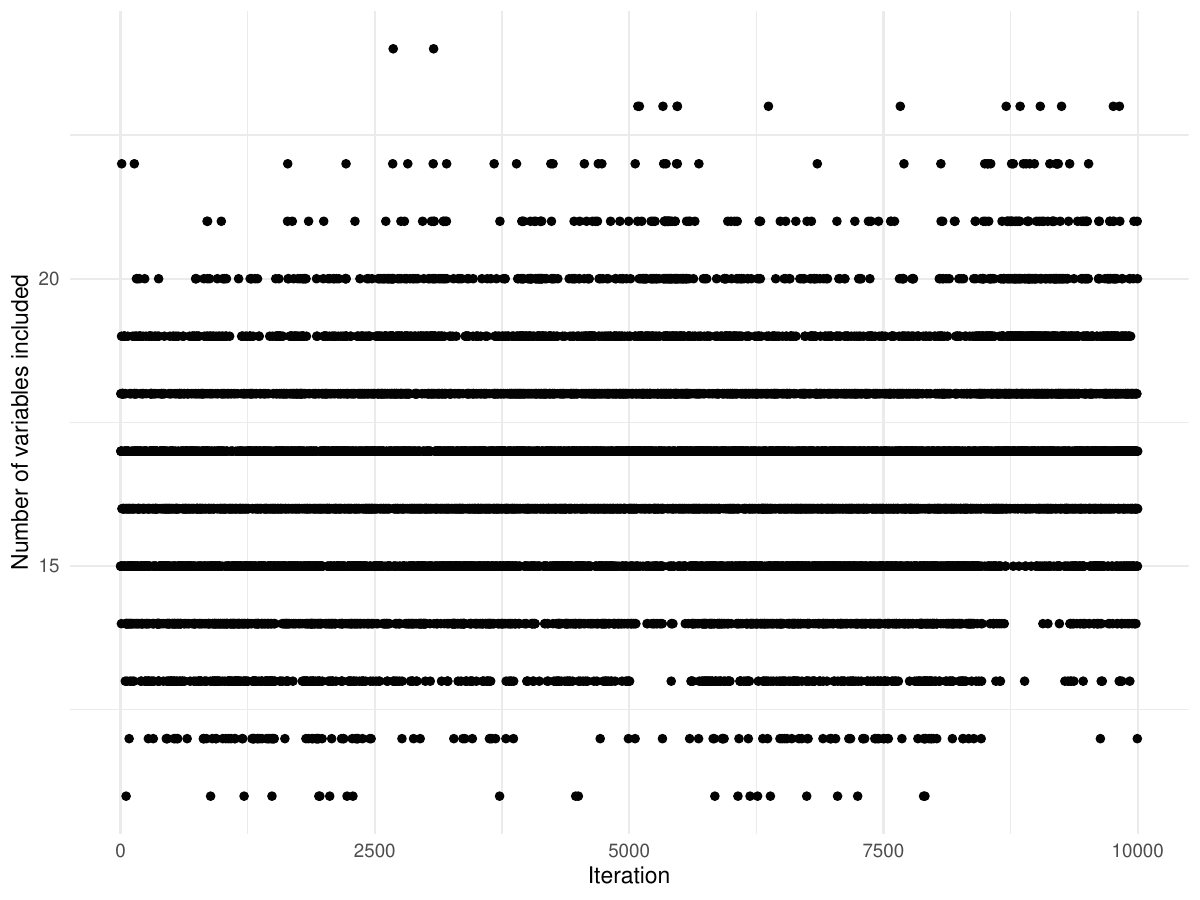}
    \includegraphics[width=0.495\linewidth]{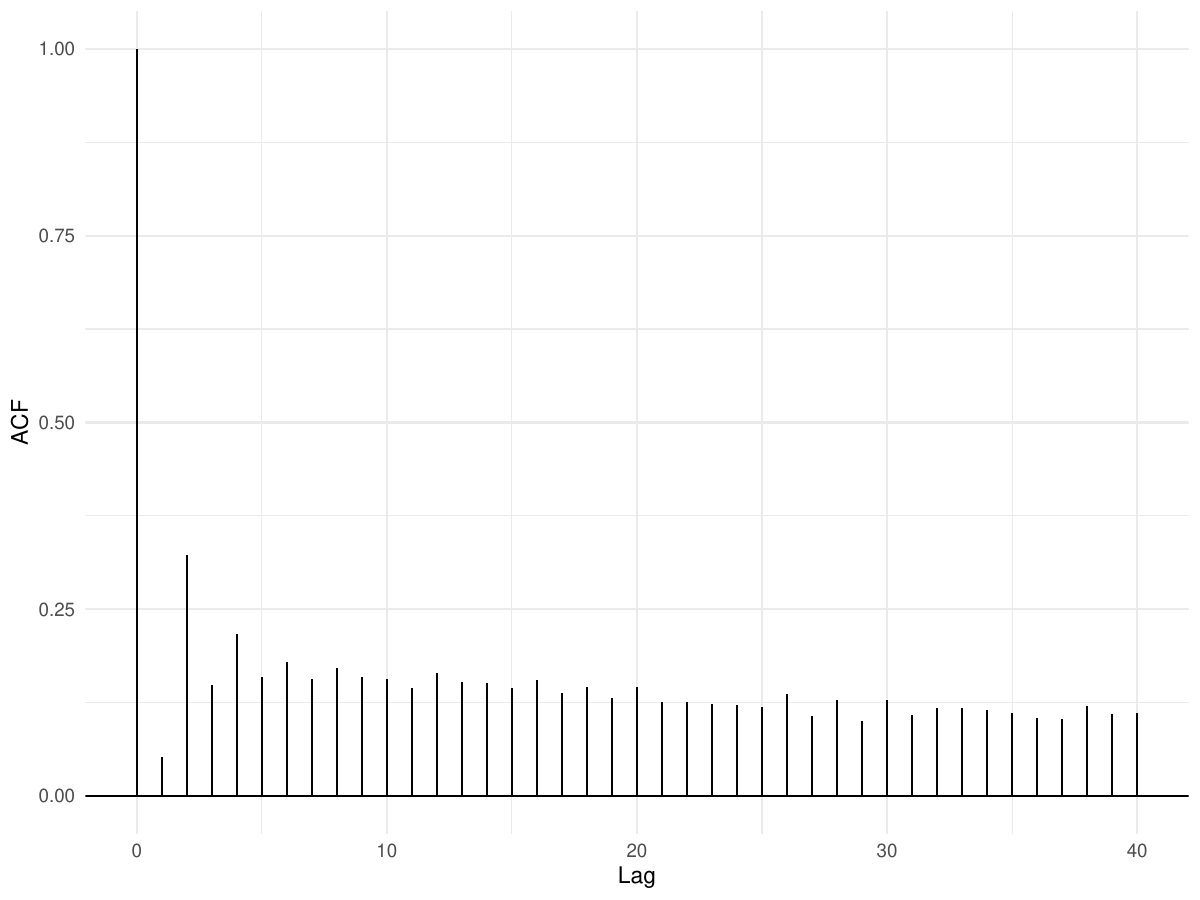}
    \caption{Convergence diagnostics: trace plot of active associations (left) and autocorrelation function (right).}
    \label{fig:DM_traceplot_autocorrelation}
\end{figure}

The corresponding convergence diagnostics when including the move step are shown in Figure~\ref{fig:DM_traceplot_autocorrelation_move}. The trace plot of the number of active associations indicates stable exploration of posterior model sizes after burn-in, while the autocorrelation function decays quickly. Together with the broader spread of non-negligible PIPs in Figure~\ref{fig:DM_PIPs_move}, these diagnostics suggest that the local move step improves local exploration of the model space while producing a more parsimonious set of selected associations. Overall, these results support the robustness of the strongest selected associations and show that the final biological conclusions are not driven by a single proposal specification.
\begin{figure}[htbp!]
    \centering
    \includegraphics[width=0.495\linewidth]{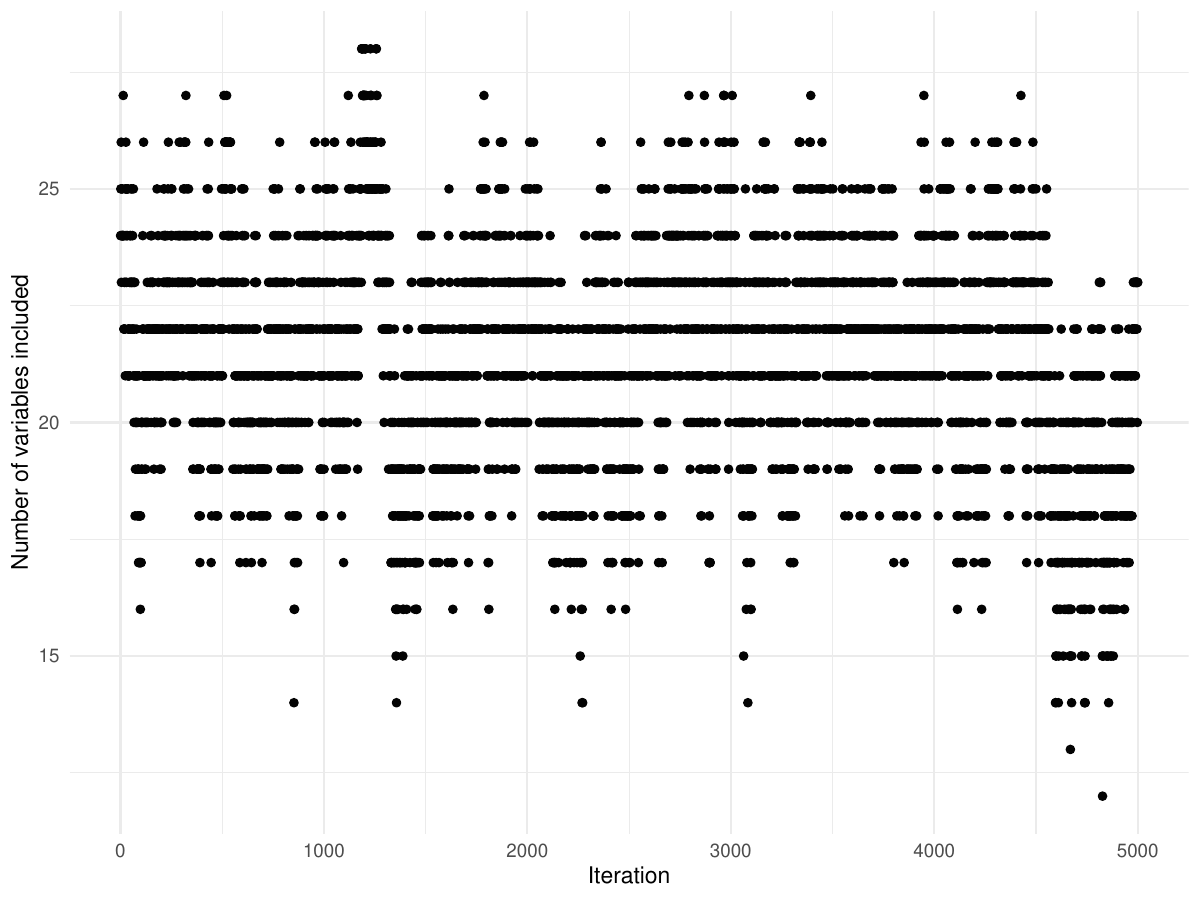}
    \includegraphics[width=0.495\linewidth]{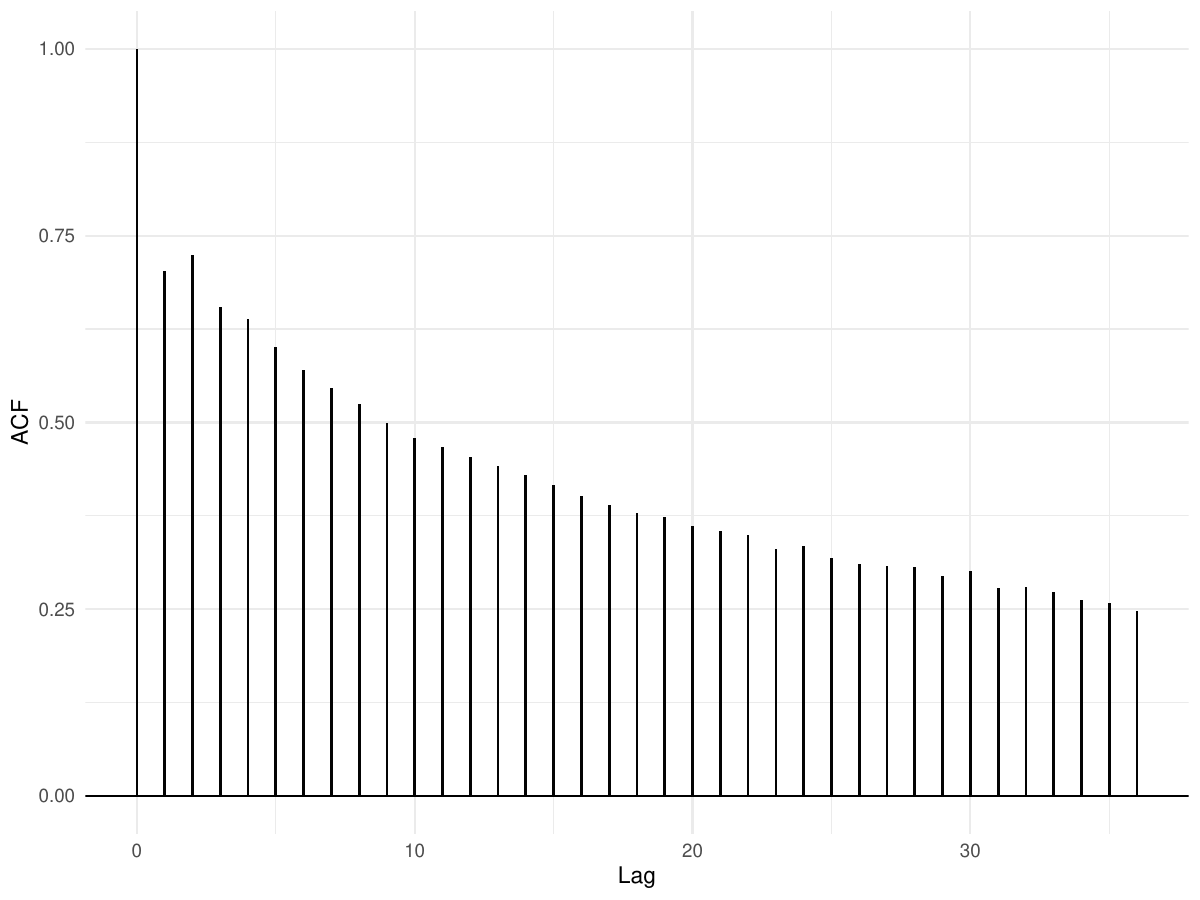}
    \caption{Convergence diagnostics for the adaptive proposal augmented by the local move step: trace plot of the number of active associations (left) and autocorrelation function (right).}
    \label{fig:DM_traceplot_autocorrelation_move}
\end{figure}

\end{document}